\documentclass[lettersize,journal]{IEEEtran}
\usepackage{amsmath,amsfonts}
\usepackage{algorithmic}
\usepackage{algorithm}
\usepackage{array}
\usepackage[caption=false,font=normalsize,labelfont=sf,textfont=sf]{subfig}
\usepackage{textcomp}
\usepackage{stfloats}
\usepackage{url}
\usepackage{verbatim}
\usepackage{graphicx}
\usepackage{cite}
\usepackage{amssymb}
\usepackage{pgffor}
\usepackage{caption}
\usepackage{amsthm}
\usepackage{color}
\usepackage{cite}
\usepackage{amsthm,amsmath,amssymb}
\usepackage{graphicx} 
\usepackage{epstopdf}

\usepackage{mathrsfs}
\newtheorem{lemma}{\bf Lemma}

\begin{document}
	
	\title{RIS-assisted Scheduling for High-Speed Railway Secure Communications}
	
	\author{Panpan Li, Yong Niu,~\IEEEmembership{Member,~IEEE}, Hao Wu,~\IEEEmembership{Member,~IEEE}, Zhu Han,~\IEEEmembership{Fellow,~IEEE},
		
		 Bo Ai,~\IEEEmembership{Fellow,~IEEE}, Ning Wang,~\IEEEmembership{Member,~IEEE}, and Zhangdui Zhong,~\IEEEmembership{Fellow,~IEEE,}
		
		\thanks{Copyright (c) 2015 IEEE. Personal use of this material is permitted. However, permission to use this material for any other purposes must be obtained from the IEEE by sending a request to pubs-permissions@ieee.org. This study was supported by the National Key Research and Development Program under Grant 2021YFB2900301; in part by National Key R\&D Program of China (2020YFB1806903); in part by the National Natural Science Foundation of China Grants 62231009, 61801016, 61725101, 61961130391, U1834210 and 61771431; in part by the State Key Laboratory of Rail Traffic Control and Safety (Contract No. RCS2021ZT009), Beijing Jiaotong University; and supported by the open research fund of National Mobile Communications Research Laboratory, Southeast University (No. 2021D09); in part by the Fundamental Research Funds for the Central Universities, China, under grant number 2022JBQY004			and 2022JBXT001; and supported by Frontiers Science Center for Smart High-speed Railway System; in part by the Fundamental Research Funds for the Central Universities 2020JBM089; in part by the Project of China Shenhua under Grant (GJNY-20-01-1); and in part by NSF CNS-2107216 and CNS-2128368; and supported by NSF CNS-2107216, CNS-2128368 and CMMI-2222810.(\emph{Corresponding author: Yong Niu.})
			
			P. Li, H. Wu, B. Ai, and Z. Zhong are with the
			State Key Laboratory of Rail Traffic Control and Safety, Beijing Jiaotong University, Beijing 100044, China (e-mail: 19111023@bjtu.edu.cn; hwu@bjtu.edu.cn; aibo@ieee.org; zhdzhong@bjtu.edu.cn).
			
			Y. Niu is with the State Key Laboratory of Rail Traffic Control and Safety, Beijing Jiaotong University, Beijing 100044, China, and also with the National Mobile Communications Research Laboratory, Southeast University, Nanjing 211189, China (e-mail: niuy11@163.com).
			
			Z. Han is with the Department of Electrical and Computer Engineering at the University of Houston, Houston, TX 77004 USA, and also with the Department of Computer Science and Engineering, Kyung Hee University, Seoul, South Korea, 446-701. (e-mail: hanzhu22@gmail.com).
		
			N. Wang is with the School of Information Engineering, Zhengzhou University, Zhengzhou 450001, China (e-mail: ienwang@zzu.edu.cn).}}

	
	\maketitle
	
	\begin{abstract}
		With the rapid development of high-speed railway systems and railway wireless communication, the application of ultra-wideband millimeter wave band is an inevitable trend. However, the millimeter wave channel has large propagation loss and is easy to be blocked. Moreover, there are many problems such as eavesdropping between the base station (BS) and the train. As an emerging technology, reconfigurable intelligent surface (RIS) can achieve the effect of passive beamforming by controlling the propagation of the incident electromagnetic wave in the desired direction. We propose a RIS-assisted scheduling scheme for scheduling interrupt flows and improving quality of service (QoS). In the propsed scheme, an RIS is deployed between the BS and multiple mobile relays (MRs). By jointly optimizing the beamforming vector and the discrete phase shift of the RIS, the constructive interference between direct link signals and reflected link signals can be achieved, and the channel capacity of eavesdroppers is guaranteed to be within a controllable range. Finally, the purpose of maximizing the number of successfully scheduled flows and satisfying their QoS requirements can be practically  realized. Extensive simulations demonstrate that the proposed scheme has superior performance regarding the number of completed flows and the system secrecy capacity over four baseline schemes in literature. 
	\end{abstract}
	
	
	\section{Introduction}
	\IEEEPARstart{I}{N}  recent years, high-speed railway (HSR) has gradually evolved from informatization to intelligence. Diversified data-intensive services have higher and higher demand for railway high-capacity communication, such as high-definition video surveillance, onboard broadband internet services, and the railway Internet of Things (IoT) service. According to current researches, the transmission rate of each train carriage is about 40 Mbps, and it may increase to 0.5$\sim$5Gbps in the future\cite{ref1}. This is a great challenge for the existing railway wireless communication system.
	
	With huge bandwidth from 30 to 300 GHz, millimeter wave (mmWave) can provide multi-gigabit communication services, such as high definition television (HDTV) and ultra-high definition video (UHDV)\cite{ref2}. Massive Multiple Input Multiple Output (MIMO) wireless communication refers to equipping base stations (BSs) and receivers with a large number of antennas. Massive MIMO has been shown to potentially improve spectral and energy efficiency \cite{ref3}.  Massive MIMO and mmWave can well support the demand of large data traffic and bandwidth-intensive applications, and solve the challenge of HSR communication. However, using a large number of active antennas in massive MIMO and mmWave implies considerable hardware costs, power consumption, and computational complexity. In addition, in a mmWave communication system, the propagation environment is not controllable. When the line of sight (LoS) link between the BS and user equipments (UEs) is blocked by the HSR carriage, a large amount of penetration loss is generated and the quality of service (QoS) is significantly reduced, which brings many design challenges.
	
	Notably, due to the broadcast nature of wireless communication, eavesdropping has always been an important security risk to users’ privacy and data security. To alleviate this problem, many studies have developed various effective authentication and encryption algorithms. However, the high-speed mobility and frequent handover of HSR bring great challenges to identity authentication. Moreover, both authentication and encryption require additional overhead and reliable third-party authentication. Fortunately, the Physical Layer Security (PLS) technology can achieve secure transmission only by using the dynamic characteristics of wireless channels to increase the capacity difference between legitimate users and illegitimate users. Moreover, the high-speed mobility of HSR makes the channel change rapidly, which brings abundant channel resources to PLS. Therefore, we implement secure flow (i.e., the traffic data transmitted between the transmitter and receiver\cite{ref4,ref5,ref6}) scheduling with PLS in this work.
	
	In recent years, reconfigurable intelligent surface (RIS) has attracted extensive attention of scholars as an innovative and revolutionary technology. RIS is a plane containing a large number of passive reflective elements, each of which can independently induce controllable amplitude and/or phase shift to the incident signals\cite{ref7}. By densely deploying RIS in a wireless communication system and subtly adjusting its parameters, it is possible to improve communication performance by increasing the expected power gain of the received signal and destructively reducing interference. As a two-dimensional implementation of metamaterials, RIS naturally has the outstanding characteristics of low cost, low complexity and easy deployment, which can well solve the challenges in intelligent HSR communication scenarios\cite{ref8,ref9}.
	
	At present, many studies have considered RIS-assisted high mobility scenarios \cite{ref10,ref11,ref12,ref13,ref14}. However, almost all of these studies focus on vehicle or UAV, and few studies focus on the RIS-assisted HSR communication. Vehicles tend to be densely distributed and have variable directions, while UAVs have more flexible three-dimensional trajectories. HSR not only moves faster, but also has a definite running direction and a regular running trajectory. These characteristics can bring both advantages to HSR communication. Using RIS to increase the coverage of mmWave BS can reduce the number of HSR handovers and improve the practicality of mmWave in HSR communication. Integrating RIS into HSR mmWave communication is challenging and very little research is currently done on this topic. Xu \textit{et al.} \cite{ref15} investigated the problem of jointly design transmission beamforming at the BS and phase shifts at the RIS for spectral efficiency maximization in RIS aided mmWave HSR networks. But he took a deep reinforcement learning approach and didn't consider the eavesdroppers. While RIS also enhances the eavesdropping signal while providing diversity gain. This is detrimental to the communication security of the HSR. In addition, the vast majority of existing researches on RIS-assisted work aim at maximizing system capacity or secrecy capacity. But our ultimate goal is to maximize the satisfaction of users, which requires optimizing the order of serving users based on their requests and system capacity. In this case, not only the coupled beamforming vector and the RIS phase shift matrix, but also the 0-1 variable indicating whether to schedule or not, are among the optimization variables. How to design the optimization variables well is important but difficult.
	
	In this paper, we consider the communication between HSR and the ground BS, introducing mmWave and RIS to improve the communication quality. Due to the penetration loss of the train carriage, we consider deploying multiple mobile relays (MRs) as communication relays on the rooftop of the train. MRs can serve passengers through the access points (APs) inside the carriage. We study a downlink RIS-assisted HSR communication system scenario where MRs request the BS to schedule a certain number of flows for them. Each flow has its own minimum throughputs requirements (i.e., minimum transfer rate requirements), referred to as its QoS requirement in our paper. Due to the quality of communication links and the existence of eavesdroppers, the flows may fail to be scheduled for not meeting the QoS requirements or too low security capacity. We propose a RIS-assisted scheduling algorithm, which aims to maximize the number of successfully scheduled flows under the constraints of QoS, secure transmission and power budget.
	
	The contributions of this paper are summarized as follows.
	\begin{itemize}
		\item{In order to improve the security capacity of the HSR communication system and meet the service requests of passengers to the greatest extent, we use the huge bandwidth of mmWave to schedule flows with different QoS requirements. We consider the communication between the BS along the track and multiple MRs on the roof. Due to the high-speed mobility of the train and the small coverage of mmWave, we deploy a RIS to provide reflection paths to enhance the received signal strength, and reduce the eavesdropping capacity of a random eavesdropper.}
		\item{We formulate a maximization problem for the number of safely scheduled flows, with the goal of optimizing the beamforming vector, the RIS phase shift matrix, and the 0-1 binary variable indicating whether to be scheduled or not, while ensuring that BS power budget, RIS phase shifts, and minimum secrecy capacity constraints. Since the optimization variables are coupled, it is difficult to optimize them at the same time. Moreover, the problem is a mixed integer non-convex nonlinear problem, so it is challenging to solve.}
		\item{We propose a low-complexity alternating optimization algorithm. The optimization problem is decomposed into three sub-problems to solve, namely beamforming, discrete phase shifts and scheduling selection optimization. The beamforming subproblem is equivalent to a Rayleigh quotient problem, which can be solved by the MRT method. The discrete phase shifts optimization subproblem employs a local search algorithm to find the optimal phase shifts. The two are alternately optimized and updated iteratively. Finally, a heuristic algorithm is developed to optimize the scheduling decisions.}
		\item {In addition, we also evaluate the proposed RIS-assisted security scheduling scheme for mmWave HSR networks through extensive simulations. The simulation results show that RIS can improve the security communication efficiency of HSR communication and expand the coverage of the cell. And our proposed algorithm has much better performance than the baseline schemes.}
	\end{itemize}
	
	The rest of the paper is organized as follows. In Section ~\ref{section: Related Work}, we summarize some related works. In Section~\ref{section: System Model}, we establish the system model, including the description of the investigated system and communication channels. In Section~\ref{section: Problem Formulation}, we formulate the problem of maximizing the number of successfully scheduled flows, and decompose it into a power allocation subproblem and a discrete phase shift optimization subproblem. Next, in Section~\ref{section: Design of Secure RIS-Assisted Wireless System}, the optimization algorithms are designed for the two subproblems respectively. Then the optimal solution is obtained by jointly optimizing the two subproblems. In Section~\ref{section: Simulation Results}, we conduct a performance evaluation and compare the proposed algorithm with other benchmark algorithms. Finally, we conclude this paper in Section~\ref{section: Conclusion}.
	
	\section{Related Work} \label{section: Related Work}
	The rapid development of railway transportation systems has greatly enriched railway wireless services and raised the demand of wireless transmission. In order to achieve high data transmission rate, the application of ultra-wideband millimeter wave is a popular trend \cite{ref16}. To promote the application of mmWave in HSR, many works have explored the propagation and fading characteristics of in HSR \cite{ref17,ref18,ref19,ref20,ref21,ref22}. This provides empirical propagation models for our research on mmWave applications. In order to ensure the stable and reliable transmission of railway mmWave signals,  have made various effective beamforming schemes. Yin \textit{et al.} \cite{ref23} proposed an mmWave-based adaptive multi-beamforming scheme. Multiple beams with different beamwidth were adopted by the BS simultaneously to improve the capacity of HSR wireless networks.  Cui \textit{et al.} \cite{ref24} proposed a novel optimal nonuniform steady mmWave beamforming scheme, to guarantee the network reliability under an interleaved redundant coverage architecture for HSR wireless systems. Whereas, Cui \textit{et al.} \cite{ref25} considered that the strong line-of-sight of the HSR propagation channel leads to high channel correlation, which makes MIMO less effective in HSR scenarios. They proposed a hybrid spatial modulation beamforming scheme operating at mmWave frequency bands for HSR wireless communication systems. The narrow beam of mmWave and the high mobility of the train make beam alignment extremely difficult. Yan \textit{et al.} \cite{ref26} took advantage of the periodicity and regularity of trains’ trajectory and proposed a fast bear alignment scheme. Gao \textit{et al.} \cite{ref27} investigated the beam tracking strategies for mmWave HSR communications, and proposed a dynamic beam tracking strategy by adjusting the beam direction and beam width jointly. In order to meet the challenges of mmWave application in HSR, Song \textit{et al.} \cite{ref28} redesigned the multiple access technology and the frame structure of OFDM and single carrier, presented train–trackside network architectures based on different MIMO techniques. The Doppler shift caused by high-speed mobility can damage HSR wireless signals. Gong \textit{et al.} \cite{ref29} conducted the modeling of the Doppler effect for mmWave in HSR communications, and designed data-aided Doppler estimation and compensation algorithms based on the new model. However, the mmWave channel has some well-known characteristics, such as high propagation loss, high penetration loss of building materials and rough-surface scattering\cite{ref30}. Yalçın \textit{et al.} \cite{ref31} used relay-assisted transmission to resistance the path loss and link blocking in mmwave communication. Other wireless access technologies, such as LTE, WiFi, etc., can be adopted for the communication inside the carriage. Therefore, we focus on the mmWave communication between the BS and MRs.
	
	The research on millimeter wave scheduling algorithm has made relatively mature progress, including time division multiple access (TDMA)\cite{ref32}, spatial time division multiple access (STDMA)\cite{ref33} and QoS awareness\cite{ref34}. Among these, TDMA is a common millimeter wave scheduling scheme, and its related studies cover different types of link scenarios, such as wireless access and backtrip networks\cite{ref35,ref36,ref37}. Therefore, in this paper, we also use TDMA scheme for millimeter wave scheduling. Millimeter wave has serious propagation loss and easy to be blocked. In order to solve the problem, beamforming in high-speed railway scenarios has been extensively studied. Gao \textit{et al.} \cite{ref38} and Zhang \textit{et al.}\cite{ref39} modeled vehicle-ground communication and used optimization methods to achieve high network throughput. Xu \textit{et al.}\cite{ref40} proposed a low-complexity iterative optimization approach of power allocation in time-varying HSR environment. Yin \textit{et al.}\cite{ref41} proposed a location-fair based mmWave stable beamforming scheme under the interlaced redundant coverage architecture to decrease the fluctuation of data rate. Xu \textit{et al.}\cite{ref42} developed an experience-driven power allocation method by leveraging multi-agent DRL, to maximize the achievable sum rate (ASR) for smart railway. However, the above work doesn't change the propagation environment, but only alleviate the dilemma of millimeter wave transmission through appropriate resource allocation and management.
	
	RIS can dynamically modify the propagation environment by adjusting its own reflection coefficient. Sun \textit{et al.}\cite{ref43} proposed a three-dimensional (3D) RIS-assisted MIMO channel model based on a 3D cylinder model. At present, there have been many related studies on the optimization problem in RIS-assisted communication. Fu \textit{et al.}\cite{ref44} and Wu \textit{et al.}\cite{ref45} studied the minimum BS transmission power optimization problem in a MISO system and a MIMO system, respectively, but they all adopted continuous RIS phase shifts. In fact, RIS's phase shift is realized by adjusting the switching state of the PIN diode. One PIN diode can realize two-phase shift. Therefore, due to the limitation of element size, continuous phase shift is unrealistic. Di \textit{et al.}\cite{ref46}  studied a downlink multi-user system, in which considered the achievable rate under the practical case where only a limited number of discrete phase shifts can be realized by a	finite-sized RIS.  Chen \textit{et al.}\cite{ref47} established a problem of maximizing the sum rate of multiple D2D links by jointly optimizing the transmission power and RIS discrete phase shifts of all links. Di \textit{et al.}\cite{ref48} proposed a hybrid beamforming scheme where the continuous digital beamforming and discrete RIS-based analog beamforming were performed at the BS and the RIS, respectively, and showed that the RIS-based system can achieve a good sum-rate performance by setting a reasonable size of RIS and a small number of discrete phase shifts. Considering the difficulty of deploying continuous phase shift in reality, we use discrete RIS phase shift in our paper.  RIS can also get a good auxiliary effect in high-speed mobile scenes. RIS was considered to be deployed in a railway wireless communication system (RWCS) for the first time to improve the anti-interference ability of RWCS communication \cite{ref49}. Chen \textit{et al.} \cite{ref50} studied the fast time-varying mmWave vehicular communication, and proposed RIS-assisted sum rate maximization algorithms for single vehicle user and multi vehicle users with imperfect CSI. Chen \textit{et al.} \cite{ref51} studied RIS-aided high-speed vehicular communication. By jointly optimizing transmission power, multi-user detection matrix, spectrum reuse and RIS reflection coefficient, they solved three major problems: spectrum sharing, imperfect CSI and QoS performance guarantee. Makarfi \textit{et al.} \cite{ref52} deployed a RIS-based access point at the source and a RIS-based relay on buildings respectively, and studied the utility of these two methods for V2V security communication. Guo \textit{et al.} \cite{ref50} studied RIS-assisted mmWave communication under imperfect CSI. At the same time, RIS phase shift, UAV active beamforming and flight trajectory are optimized to ensure the maximum security capacity with multiple eavesdropping. Ren \textit{et al.} \cite{ref51} optimized the similar indicators as \cite{ref53}, but they were committed to minimizing the UAV energy consumption.While HST is very different from vehicles and UAV. HST tends to be located in suburban, mountainous and other fields, and has fixed routes and smooth speed. What’s more, there are almost no two HSTs at the same time. Based on this, HST communication is quite different from the Internet of Vehicles and UAV communication. Therefore, it is of great significance to study RIS-assisted HSR communication. But so far, only Xu \textit{et al.}\cite{ref15} have studied the performance of RIS in HSR communication. However, in these studies, the existence of eavesdroppers and the scheduling of traffic flows are not considered.
	
	Due to the broadcast characteristics of wireless communication, it is very vulnerable to be eavesdropped and attacked\cite{ref55}. In recent years, PLS (Physical Layer Security) has received extensive attention \cite{ref56,ref57,ref58}. PLS uses the dynamic characteristics of wireless channel to realize secure communication without complicated encryption and decryption process. At the same time, the advantage of RIS is to change the channel propagation environment. Therefore, RIS can be used to enhance the performance of PLS. Zhang \textit{et al.} \cite{ref59} derived the key indicators of RIS-aided communication, such as the secrecy outage probability, the probability of nonzero secrecy capacity, and proved that RIS can provide good performance for PLS. Zhang \textit{et al.} \cite{ref60} studied the impact of RIS on secure communication in the four cases of internal eavesdropping, external eavesdropping and with/without channel state information (CSI) of eavesdroppers. Makarfi \textit{et al.} \cite{ref61} and Gu \textit{et al.} \cite{ref62} respectively studied the influence of the location of RIS and Eve on the security performance. Dong \textit{et al.} \cite{ref63} proposed a new design of active RIS, which can amplify the signal strength and achieve better security performance gain than passive RIS. As a key infrastructure, the security of high-speed railway wireless communication is very important. Xu \textit{et al.} \cite{ref64} described the security attacks faced by HSR communication and the practical dilemma of the application of traditional security technologies in high-speed mobile scenarios. Therefore, we hope to take advantage of the rapid change of wireless channels caused by the high-speed mobility of HSR, and use RIS-assisted PLS to achieve security and effective traffic flow scheduling.

	\begin{figure}[t]
	\centering
	\includegraphics[width=3.2in,height=2.3in]{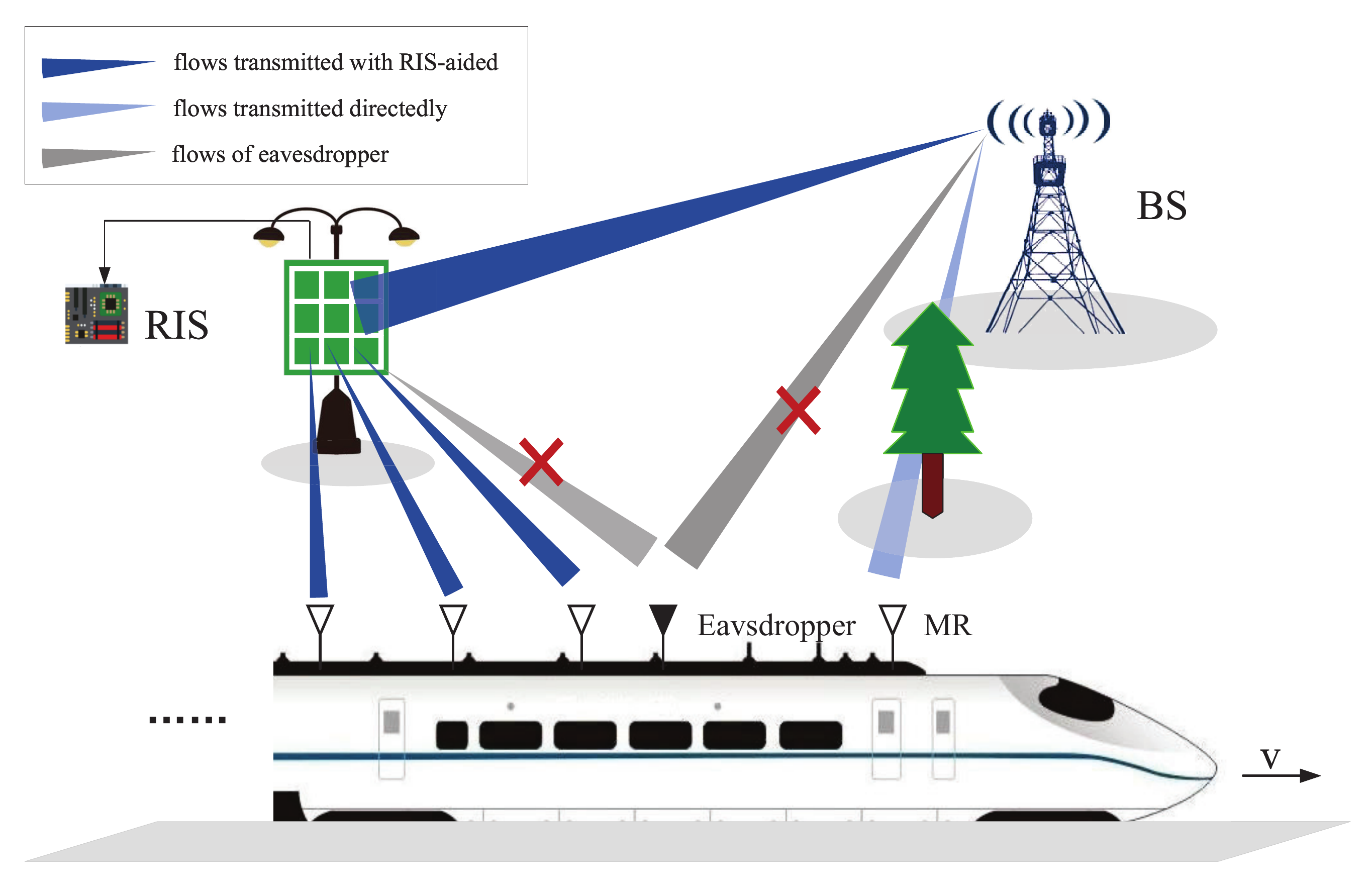}
	\caption{The mmWave HSR communication system.}
	\label{Fig:System model}
	\label{fig_1}
\end{figure}
	
	\section{System Model} \label{section: System Model}
	\subsection{System Description}
	
We consider a single cell scenario of the mmWave HSR communication shown in Fig. \ref{Fig:System model}. A BS equipped with $M$ antennas is fixed beyond the railway track. There are $N$ MRs deployed on the rooftop of the train. An RIS is deployed to assist the communication between the BS and MRs, which is composed of a controller and a great quantity of programmable elements. It reflects signals that impinge on the surface and emmits them in the disired beamform. Specifically, the controller  tunes the voltage-controlled PIN diodes of each RIS element in real time by changing their switching state, and then the amplitude and phase of the incident wave will be adjusted to expected value. By using this reconfigurable property and combining with channel estimation, the legitimate channel capacity can be improved, and the eavesdropping channel can be suppressed. The BS and RIS are located in fixed positions on the same side of the track, and all MRs are within the coverage of the BS and RIS assistance. There are two kinds of links in the system, the one is the direct transmission links from the BS to MRs, as shown in the light blue link in Fig. \ref{Fig:System model}, and the other one is the RIS-assisted reflection links, as shown in the dark blue link in Fig. \ref{Fig:System model}. There is also an eavesdropper in the system that eavesdrops illegally on signals messages, as shown in the gray link in Fig. \ref{Fig:System model}.
	
In our investigated system, the RIS is a uniform plannar array connsisting of $L$ elements. The range of phase shifts of each element is constrained in $[0,2\pi]$. Considering the realities of hardware configuration, we just take finite discrete values with equal quantization intervals with the valid range. We assume that the number of quantization bits is $e$, and then the phase shifts of each element can be represented as  $\phi_l=\frac{2m_l\pi}{2^e-1}$, where $l=1,2,\dots,L$, $m_l=\left \{ 0,1,\cdots,2^e-1 \right \}$. In our system, time is divided into a series of non-overlapping superframes, each of which consists of the scheduling phase and the transmission phase. The scheduling phase is the duration of collecting requested services and MRs' QoS requirements. There are $F$ flows need to be scheduled between the BS and MRs, and each flow can be directly received from the BS to MRs or assisted with RIS's reflection.  Since the moving direction and velocity of the train can be known a priori and is predictable, the doppler frequency shifts  of the signals received by the MRs are assumed to be known and can be eliminated with some existing technologies.

At present, there are many studies on CSI estimation of RIS-assisted channel. Various practical approaches have been proposed. For example, Nadeem \textit{et al.}\cite{ref65} proposed a minimum mean suqared erroe based channel estimation propocol, and  Wang \textit{et al.}\cite{ref66} established a novel three-phase pilot-based channel estimation freamwork. Tang \textit{et al.} [67] developed free-space path loss models for RIS-assisted wireless communications for different scenarios by studying the physics and electromagnetic nature of RISs. It pointed out that the free space path loss of BS-RIS-User channel is proportional to $(d_{R,2} d_{n,3} )^2$ in RIS assisted far-field communication. Chen \textit{et al.} [47] used Rice fading as the small-scale fading model. Since our research focuses on resource allocation, channel estimation is beyond our research content. Therefore, in our paper, we assume that we have perfect channel state information, and channel modeling refers to the empirical models of \cite{ref67} and \cite{ref47}.

	\subsection{Pass Loss Model}
	We use $d_{n,1}$ and $d_{n,3}$ to denote the distance from MR $n$  $(n=1,2,...N)$ to the BS and RIS, respectively. And the distance between the BS and RIS is denoted by  $d_{R,2}$. Then the pass loss of the LoS link from the BS to the MRs, RIS, and from the RIS to MRs\cite{ref67}, respectively, are
	
	\begin{equation}
		\label{deqn_ex1a}
		L(d_{n,1})=Cd_{n,1}^{-\alpha_1},
	\end{equation}

	\begin{equation}
		\label{deqn_ex1a}
		L(d_{R,2})=Cd_{R,2}^{-\alpha_2},
	\end{equation}

	\begin{equation}
		\label{deqn_ex1a}
		L(d_{n,3})=Cd_{n,3}^{-\alpha_2},
	\end{equation}
	
	\noindent where $C$ is the path loss intercept, $\alpha_1$ is the path loss exponent in the LoS case, and $\alpha_2$ is the path loss exponent in the NLoS case.
	
	\subsection{Small-Scale Fading}
	In order to construct a practical RIS framework, We assume that the small-scale fading of each link follows the Rican distribution with a $\beta_l$ Rican factor. The channel correlation between RIS elements may exist because the electrical size of RIS's reflecting elements is between $\lambda/8$ and $\lambda/4$ in principle, where $\lambda$ is a wavelength of the signal \cite{ref68}. The small-scale fading matrices between the BS and MRs are defined as
	
	\begin{equation}
		\label{deqn_ex1a}
		\boldsymbol Q_{n,1}=[q_1^{n,1}, q_2^{n,1},\cdots q_M^{n,1}],
	\end{equation}
	
	\noindent where $\boldsymbol Q_{n,1}$ is a $1\times M$ matrix. The elements in $\boldsymbol Q_{n,1}$ are represented by Rican factor $\beta_l$ and the random variables $q_m^{ray}$ follow the Rayleigh distribution\cite{ref47}, i.e.,
	
	\begin{equation}
		\label{deqn_ex1a}
		q_m^{n,1}=\sqrt{\frac{\beta_l}{\beta_l+1}}+\sqrt{\frac{1}{\beta_l+1}}q_m^{ray}.
	\end{equation}
	
	Similarly, we use the matrice $\boldsymbol Q_{R,2}\in \mathbb{C}^{L\times M}$ to denote the small-scale fading of the link from the BS to the RIS, and use matrice $\boldsymbol Q_{n,3}\in \mathbb{C}^{1\times L}$ to denote the small-scale fading of the link from the RIS to the MRs. They are, respectively, defined as
	
	\begin{equation}
		\label{deqn_ex1a}
		\boldsymbol Q_{R,2}=\begin{bmatrix}
			q_{1,1}^{R,2} &\cdots   &q_{1,M}^{R,2} \\ 
			\vdots & \ddots  &\vdots  \\ 
			q_{L,1}^{R,2}&  \cdots & q_{L,M}^{R,2}
		\end{bmatrix},
	\end{equation}
	
	\begin{equation}
		\label{deqn_ex1a}
		\boldsymbol Q_{n,3}=\begin{bmatrix}
			q_1^{n,3},&q_2^{n,3},&\cdots&q_L^{n,3}     
		\end{bmatrix}.
	\end{equation}
	
	Therefore, the channel coefficients of the BS-MRs links, the BS-RIS link, and the RIS-MRs links can be denoted as $\boldsymbol d_n\in \mathbb{C}^{1\times M}$, $\boldsymbol G\in \mathbb{C}^{L\times M}$, $\boldsymbol h_n\in \mathbb{C}^{1\times L}$,
	
	\begin{equation}
		\label{deqn_ex1a}
		\boldsymbol d_n=\sqrt{L(d_{n,1})}\boldsymbol Q_{n,1},
	\end{equation}
	
	\begin{equation}
		\label{deqn_ex1a}
		\boldsymbol G=\sqrt{L(d_{R,2})}\boldsymbol Q_{R,2},
	\end{equation}
	
	\begin{equation}
		\label{deqn_ex1a}
		\boldsymbol h_n=\sqrt{L(d_{n,3})}\boldsymbol Q_{n,3}.
	\end{equation}
	
	\subsection{Achievable Rate and Secrecy Rate}
	In the DL transmission, the complex baseband transmitted signal at the BS can be expressed as
	
	\begin{equation}
		x_n= \boldsymbol w_ns_n,
	\end{equation}
	
	\noindent where $\boldsymbol w_n=\left [ w_n^1,w_n^2,\cdots,w_n^M \right ]^T$ is the beamforming vector, satisfing the constraint of $\left \| \boldsymbol w_n \right \|^2\leq P_{max}$, and  $s_n$ $(n = 1,\cdots,N)$ are i.i.d. random variables (RVs) with zero mean and unit variance, denoting the informationbearing symbols of users.
	
	For MR $n$, a binary variable $a_n^k$ is defined to indicate whether its request is scheduled in the $k$th time slot. If it is scheduled, $a_n^k=1$; otherwise, $a_n^k=0$.	Then the signal received by MR $n$ in the $k$th time slot can be obtained
	
	\begin{equation}
		y_n^k=a_n^k(\boldsymbol d_n+\boldsymbol h_n\boldsymbol{\Phi G})\boldsymbol w_ns_n+N_0,
	\end{equation}
	
	\noindent
	\noindent
	where $\boldsymbol{\Phi} \triangleq\displaystyle diag\begin{bmatrix}e^{j\phi_1},&e^{j\phi_2},&\cdots,&e^{j\phi_L}\end{bmatrix}$ is a diagonal matrix considering the effective phase shifts introduced by all elements of the RIS.  $N_0$ denotes the additive white Gaussian noise, which is modeled as a realization of a zero-mean complex circularly symmetric Gaussian variable with variance $\sigma _N^2$. Accordingly, the SNR of the signal received in the $k$th time slot of MR $n$ is
	
	\begin{equation}
		\Gamma_n^k=a_n^k\frac{\left | (\boldsymbol d_n+\boldsymbol h_n\boldsymbol\Phi\boldsymbol G)\boldsymbol w_n \right |^2}{\sigma_N^2}.
	\end{equation}
	
	According to the Shannon’s capacity formula,  the corresponding achievable rate received by MR $n$  in the $k$th time slot is given as
	\begin{equation}
		R_n^k=log_2\left(a_n^k\frac{\left|\left( \boldsymbol d_n+\boldsymbol h_n\boldsymbol \Phi \boldsymbol G \right )\omega_n\right|^2}{\sigma_N^2} \right ).
	\end{equation}
	
	Then the throughput achieved by MR $n$ in a superframe can be obtained as
	
	\begin{equation}
		R_n=\frac{\sum_{k=1}^{K}R_n^k\cdot \bigtriangleup T}{T_s+K\cdot \bigtriangleup T},
	\end{equation}
	
	\noindent
	where $T_s$ is the duration of the scheduling phase and $\bigtriangleup T$ is the duration of every transmission phase.
	
	Similarly, in the $k$th time slot, the signal eavesdropped by the eavesdropper is
	
	\begin{equation}
		\begin{aligned}
			y_e^k ={} & a_n^k(\boldsymbol d_e+\boldsymbol h_e\boldsymbol{\Phi G})\boldsymbol w_ns_n+N_0\\
			={} & a_n^k(\sqrt{L(d_{e,1})}\boldsymbol Q_{e,1} \\
			& +\sqrt{L(d_{R,2})L(d_{e,3})}\boldsymbol Q_{e,3}\boldsymbol \Phi\boldsymbol Q_{R,2})\boldsymbol w_ns_n+N_0,
		\end{aligned}
	\end{equation}
	
	\noindent where $\boldsymbol d_{e,1}$ represents the LoS distance from the BS to the eavesdropper, $\boldsymbol Q_{e,1}\in \mathbb{C}^{1\times M}$ represents the small-scale fading from the BS to the eavesdropper. Then the SNR of the signal received by the eavesdropper in the $k$th time slot is
	
	\begin{equation}
		\Gamma_e^k=a_n^k\frac{\left | (\boldsymbol d_e+\boldsymbol h_e\boldsymbol\Phi\boldsymbol G)\boldsymbol w_n \right |^2}{\sigma_N^2}.
	\end{equation}
	
	The eavesdropping channel capacity in the $k$th time slot can be obtained on the basis of the Shannon’s capacity formula
	
	\begin{equation}
		R_e^k=log_2\left(a_n^k\frac{\left|\left( \boldsymbol d_e+\boldsymbol h_e\boldsymbol \Phi \boldsymbol G \right )\omega_n\right|^2}{\sigma_N^2} \right ).
	\end{equation}
	
	Then the throughput achieved by the eavesdropper in a superframe can be expressed as
	
	\begin{equation}
		R_e=\frac{\sum_{k=1}^{K}R_e^k\cdot \bigtriangleup T}{T_s+K\cdot \bigtriangleup T}.
	\end{equation}
	
	The secrecy capacity is an important physical layer security performance evalution metrics which can be used to determine the feasibility of the schedule scheme. For MR $n$, its security capacity is given by
	
	\begin{equation}
		C_n=\left [ R_n-R_e \right ]^+,
	\end{equation}
	
	\noindent where $\left [ x \right ]^+\triangleq max(x,0)$.
	
	There are $F$ flows come from $N$ MRs ($F\leq N$). Each flow has its own QoS requirement $q_f$, which is the minimum throughput required for its actual transmission. $\delta_f$ is defined to express whether flow $f$ is successfully scheduled, if so, $\delta_f=1$; otherwise, $\delta_f=0$. The value of $\delta_f$ is related to the quality of service requirement $q_f$ and the actual throughput of the link.
	
	\section{Problem Formulation} \label{section: Problem Formulation}
	In this section, we first formulate our optimization problem based on the above system model, and then decompose the complex optimization problem into three sub-problems in order to obtain a sub-optimal solution efficiently.
	
	\subsection{Scheduled Flows Maximization Problem Formulation}
	
	we consider the joint optimization of transmission scheduling, power controlling and RIS phase shifts, to maximize the number of scheduled flows successfully.
	
	
	\begin{equation}
		\textup{max}\sum_{i=1}^{F}\delta _f=\sum_{n=1}^{N}\left ( \left (\sum_{k=1}^{K}a_n^k  \right )\neq 0 \right ),
	\end{equation}
	
	where $\sum_{k=1}^{K}a_n^k$ represents the sum of $a_n^k$ in $K$ time slots for MR $n$.  $(\sum_{k=1}^{K}a_n^k)\neq0$ is a judgment statement. If $(\sum_{k=1}^{K}a_n^k)\neq0$ is established, then its value is $1$, which proves that the requested flow f of MR $n$ is successfully scheduled, otherwise it equals to $0$. Then all N MRs are summed to obtain the total number of successfully scheduled flows.
	
	Due to the existence of the eavesdropper, the confidentiality of transmission must be guaranteed. RIS can not only improve the data transmission rate of legitimate MRs, but also reduce the data rate of the eavesdropper, so as to increase the difference between the two rates. We guarantee that the security capacity is at least $10\%$ of the legal channel capacity. The security capacity constraint of MR $n$ is given by the following formula
	
	\begin{equation}
		C_n\geq0.1\cdot R_n,n\in \boldsymbol N.
	\end{equation}
	
	Moreover, since there are $K$ time slots available for scheduling in each superframe, the total number of time slots scheduled for all users should not be greater than $K$, which can be expressed by the following formula
	
	\begin{equation}
		\sum_{n=1}^{N}\sum_{k=1}^{K}a_n^k\leq K.
	\end{equation}
	
	Therefore, the problem of maximizing the number of scheduled flows in limited time slots can be expressed as the following optimization problem
	
	\begin{equation}
		\begin{aligned}
			\hspace{0.3cm}&\underset{\boldsymbol w_n,\boldsymbol \Phi,a_n^k}{\textup{max}}\sum_{f\in F}^{}{\delta_f}\\
			s.t.\hspace{0.3cm}&(a) \left \| \boldsymbol w_n \right\|_2^2\leq P_{max},\\
			&(b)\left [ \boldsymbol \Phi \right ]_{l,l}=e^{j\phi_l},\phi_l=\frac{2m_l\pi}{2^e-1},\\
			&\hspace{0.6cm}l=1,\cdots ,L,m_l\in\left \{ 0,1,\cdots,2^e-1 \right \},\\
			&(c)R_i^k=log_2(1+a_n^k\frac{\left | (\boldsymbol d_i+\boldsymbol h_i\boldsymbol \Phi \boldsymbol G)\boldsymbol w_n \right |^2}{\sigma_N^2})\\
			&\hspace{0.6cm}R_i=\frac{\sum_{k=1}^{K}R_i^k\cdot \Delta T}{T_s+K\cdot \Delta T},i=\left \{ e,n \right \},n=\left \{ 1,\cdots N \right \},\\
			&(d)\hspace{0.1cm}C_n=R_n-R_e>0.1\cdot R_n,n=1,\cdots ,N,\\
			&(e)\hspace{0.1cm}\sum_{n=1}^{N}\sum_{k=1}^{K}a_n^k\leq K,\\
			&(f)\hspace{0.1cm}\delta_f=\left\{\begin{matrix}
				1, & \frac{q_f\times (T_s+K\cdot\bigtriangleup T)}{R_n\times\bigtriangleup T}\geq T_{rem}\bigcap R_n\geq q_f, \\ 
				0, & otherwiese,
			\end{matrix}\right.
		\end{aligned}
	\end{equation}
	
	where constraint ($a$) indicates that the maximum BS transmission power is limited to $P_{max}$; constraint ($b$) is the phase shift of RIS elements; constraint ($c$) is the legal/illegal capacity; constraint ($d$) is the security capacity limit of the link; constraint ($e$) indicates that the total number of time slots scheduled by all users in a superframe cannot be greater than the number of time slots $K$ of the superframe;constraint ($f$) defines the value of $\delta_f$, $T_{rem}$ is the time slots remained, only when the current system capacity meets the QoS requirements of flow $f$ and the flow $f$ can be scheduled within the remaining time slots, $\delta_f=1$, otherwise $\delta_f=0$.

	\subsection{Problem Decomposition}
	For the joint optimization problem, since only one flow is considered to be scheduled at the same time, we can firstly optimize each flow to find the corresponding optimal beamforming vector and phase shift matrix, and then schedule them in the priority order calculated based on the number of required slots. Whereas, it worths pointing out that the joint optimization of the beamforming vector and phase shift matrix for each flow is coupled, and so it is difficult to optimize them at the same time. Therefore, we consider optimizing them individually. Specifically, the local search algorithm is used to traverse all possible phase shifts firstly. Next, for each phase shift selection, the BS transmit power beamforming vector is designed to maximize the security capacity of the legitimate channel while minimize the channel capacity of the eavesdropper. Finally, the phase shift matrix with the largest security capacity and the corresponding beamforming vector will be selected.

	\emph{1) Transmit Beamforming Design}: The sub-problem is allocating the appropriate transmission power within the power constraint to maximize the capacity of all MRs. When the phase shift variable $\boldsymbol{\Phi}$ is fixed, the problem with respect to the beamforming vector $\boldsymbol w_n$ in (24) can be written as
	
	\begin{equation}
		\begin{aligned}
			\hspace{0.3cm}&\underset{\boldsymbol w_n}{\textup{max}}\sum_{n=1}^{N}{C_n}\\
			s.t.\hspace{0.3cm}&\left \| \boldsymbol w_n \right\|_2^2\leq P_{max},
		\end{aligned}
	\end{equation}
	
	\noindent where $C_n=[R_n-R_e]^+$. The operator $[\cdot]^+$ has no effect on the optimal solution, and so it can be omitted for the sake of simplification in the following paragraphs.
	
	\emph{2) Discrete Phase Shifts Optimize}: The sub-problem is to select the optimal phase shift in the finite discrete phase shifts on the basis of satisfing the security capacity constraint, so as to maximize the MRs' channel capacity. When the beamforming vector $\boldsymbol w_n$ is fixed, the optimizing problem about the phase shift matrix $\boldsymbol \Phi$ in ${(24)}$ can be written as
	
	\begin{equation}
		\begin{aligned}
			\hspace{0.3cm}&\underset{\boldsymbol \Phi}{\textup{max}}\sum_{n=1}^{N}{C_n}\\
			s.t.\hspace{0.3cm}&\left [ \boldsymbol \Phi \right ]_{l,l}=e^{j\phi_l},\phi_l=\frac{2m_l\pi}{2^e-1},\\
			&\hspace{0.6cm}l=1,\cdots ,L,\\
			&\hspace{0.6cm}m_l\in\left \{ 0,1,\cdots,2^e-1 \right \}.
		\end{aligned}
	\end{equation}
	
	\emph{3) Scheduling Sequence Organize}: The sub-problem is to sort the flows to be scheduled according to the known maximum security capacity of each user, after optimizing the transmission beamforming vector and phase shift matrix. At this point, calculate the number of required slots scheduling each flow on the basis of their minimum throughput requirements. In addition, the flows are sorted in order of the number of slots from least to most. Then the sequential flows are scheduled to meet the goal of maximizing the total number of successfully scheduled flows.

	\section{Design of Secure RIS-Assisted Wireless System} \label{section: Design of Secure RIS-Assisted Wireless System}
	
	In this section, we propose a RIS-assisted scheduling scheme. Each flow can be transmitted directly and  reflected by RIS simultaneously, and all RIS elements have a limited number of discrete phase shifts. Therefore, the proposed scheme should first determine the optimial transmission beamforming vector and the phase shift matrix, and then plan all related links that need to be scheduled. Ultimately the goal of maximizing the number of scheduled flows can be achieved.
	
	\subsection{Transmit Beamforming Subproblem Algorithm Design}
	
	We first study the optimization of the beamforming vector $\boldsymbol w_n$ with the fixed phase shift matrix $\boldsymbol \Phi$. According to (14), (18) and (20), (25) can be deformed. The design problem of the deformed beamforming is given by the following formula\cite{ref69}
	
	\begin{equation}
		\begin{aligned}
			\hspace{0.3cm}&\underset{\boldsymbol w_n}{\textup{max}}\sum_{n=1}^{N}{\frac{1+\frac{1}{\sigma_N^2}\left | (\boldsymbol d_n+\boldsymbol h_n^{\boldsymbol H}\boldsymbol \Phi\boldsymbol G)\boldsymbol w_n) \right |^2}{1+\frac{1}{\sigma_N^2}\left | (\boldsymbol d_e+\boldsymbol h_e^{\boldsymbol H}\boldsymbol \Phi\boldsymbol G)\boldsymbol w_n) \right |^2}}\\
			s.t.\hspace{0.3cm}&\left \| \boldsymbol w_n \right\|_2^2\leq P_{max}.
		\end{aligned}
	\end{equation}
	
	The optimal solution is given by the following lemma.
	\begin{lemma} 
		Given the phase shift matrix $\boldsymbol \Phi$ of RIS, the optimal solution of the beamforming vector $\boldsymbol w_n$ is given by the following formula
	
	\begin{equation}
		\boldsymbol w_n^*=\sqrt{P_{max}}\lambda_{max}(\boldsymbol X_e^{-1}\boldsymbol X_n),
	\end{equation}
	
	\noindent where
	
	\begin{equation}
		\boldsymbol X_i=\boldsymbol I_M+\frac{P_{max}}{\sigma_N^2}[(\boldsymbol h_i\boldsymbol \Phi\boldsymbol G+\boldsymbol d_i)^H(\boldsymbol h_i\boldsymbol \Phi\boldsymbol G+\boldsymbol d_i)],i\in\left \{ n,e \right \}.
	\end{equation}
	\end{lemma}
	
		\renewcommand{\qedsymbol}%
		{\rule{1ex}{1.5ex}}
		\begin{proof}
			It was shown that \cite{ref70}, in MISO channels, the beamforming vector at the transmitter is optimal to allocate all transmission power to the legal receiver, that is: $\left \| \boldsymbol w_n^* \right \|^2=P_{max}$.  Therefore, the numerator and denominator of the objective function (27) can be rewritten as
	
		\begin{equation}
			\begin{aligned}
				&1+\frac{P_{max}}{\sigma_N^2}\left | (\boldsymbol d_n+\boldsymbol h_n\boldsymbol \Phi\boldsymbol G)\boldsymbol w_n \right |^2\\
				&=\widetilde{\boldsymbol w_n}^H\widetilde{\boldsymbol w_n}\\
				&\hspace{0.4cm}+\frac{P_{max}}{\sigma_N^2}\widetilde{\boldsymbol w_n}^H\left [ (\boldsymbol h_i\boldsymbol \Phi\boldsymbol G+\boldsymbol d_i)^H(\boldsymbol h_i\boldsymbol \Phi\boldsymbol G+\boldsymbol d_i) \right ]\widetilde{\boldsymbol w_n}\\
				&\triangleq\displaystyle \widetilde{\boldsymbol w_n}^H\boldsymbol X_i\widetilde{\boldsymbol w_n},
			\end{aligned}
		\end{equation}
	
		\noindent where  $\widetilde{\boldsymbol w_n}=\boldsymbol w_n/\sqrt{P_{max}}$ is a unit vector. Substituting (30) into the objective function in (27), we can get
	
		\begin{equation}
			\begin{aligned}
				\hspace{0.3cm}&\underset{\boldsymbol w_n}{\textup{max}}\sum_{n=1}^{N}\frac{\widetilde{\boldsymbol w_n}^H\boldsymbol X_i\widetilde{\boldsymbol w_n}}{\widetilde{\boldsymbol w_n}^H\boldsymbol X_e\widetilde{\boldsymbol w_n}}\\
				s.t.\hspace{0.3cm}&\left \| \boldsymbol w_n \right\|_2^2\leq P_{max},\forall n.
			\end{aligned}
		\end{equation}
	
		In this way, we transform the power optimization subproblem (27) into a 	generalized eigenvalue problem, and the optimal solution is given by (28).
		\end{proof}
	
	\subsection{Discrete Phase Shift Optimization Subproblem Algorithm Design}
	When the transmitting beamforming vector is fixed and scheduling subproblem is not considered, the power constraint and scheduling constraint are removed, and the optimization objective is shown in (26). The objective function and constraint on $\boldsymbol \Phi$ are still nonconvex. In addition, $\boldsymbol \Phi$ contains a series of discrete variables, and the range available for each phase shift depends on RIS quantization bits $e$. Considering the complexity, we use a local search method as shown in Algorithm \ref{Algorithm:1} to solve this problem. Specifically, for each element, we traverse all possible phase shifts while keeping the other $L-1$ phase shifts unchanged. Then, the corresponding optimal beamforming vector is calculated by (28), and the optimal value is selected according to the target of maximizing the security rate. Then, this optimal solution $\phi_l^*$ is used as a new value of $\boldsymbol \Phi$ to optimize other phase shifts, until $\boldsymbol \Phi$ is fully optimized.

\begin{algorithm}[H]
	\caption{Local Search for Phase Shift}\label{alg:alg1}
	\begin{algorithmic}[1]
		\REQUIRE the number of quantization bits: $e$
		\ENSURE \boldmath$\Phi^*,\omega^*$
		\FOR{$l=1:L$}
		\STATE Assign all possible values to $\phi_l$, and use (28) to find the corresponding optimal beamforming vector $\omega$; select	the value maximizing the secrecy rate,\\
		denoted as $\phi_l^*$; $\phi_l^*=\phi_l$;
		\ENDFOR
	\end{algorithmic}
	\label{Algorithm:1}
	\label{alg1}
\end{algorithm}	
	
\begin{algorithm}[H]
	\caption{The RIS-Assisted Scheduling Algorithm}\label{alg:alg1}
	\begin{algorithmic}[1]
		\REQUIRE $F_1=F$,$F_2=\emptyset$,$N_F=0$,$\delta_f=0,\forall f\in F$,\newline
		$T=0$,$q_f,R_n,\forall n\in [1,2,...,N],\Delta T,T_s,K$
		\ENSURE $N_F$
		\STATE {Calculate the priority values of the flows in} $F_1$
		\STATE {Sort the links in} $F_1$ {in decreasing order by priority values}
		\FOR {flow $f\in F_1$}
		\STATE $t_f=\qquad \frac{q_f\times(T_s+K\times\Delta T)}{R_n\times\Delta T}$\\
		\IF{$T+t_f\leq K$} 
		\STATE $\delta_f=1,F_2=F_2\cup f,N_F=N_F+1$ 
		\ELSE 
		\STATE
		\ENDIF 
		\ENDFOR
	\end{algorithmic}
	\label{Algorithm:2}
	\label{alg1}
\end{algorithm}
	
	\subsection{Scheduling Sequence Optimization Subproblem Algorithm Design}
	In this subsection, we consider the resource scheduling optimizing problem in the RIS-assisted system. Based on the above two subproblem optimization algorithms, we obtain the optimal transmission beamforming vector, the phase shift matrix, and the maximum security capacity. Then, since we only serve one flow once a time, we consider the scheduling order of all flows. Define the parameter $\widetilde{\boldsymbol \omega_f}$ to determine the priority of the flow $f$, which is the reciprocal of the number of slots that it needs in a frame to meet its QoS requirement. For flow $f$ requested by MR $n$, the priority value $\widetilde{\boldsymbol \omega_f}$ can be expressed as
	
	\begin{equation}
		\widetilde{\boldsymbol \omega_f}=\frac{R_n\cdot\Delta T}{q_f\cdot(T_s+K\cdot\Delta T)}.	
	\end{equation}
	
	The less time it takes to complete a flow, the more flows can be completed in a given amount of time slots. In order to complete as many users' requests as possible, we sort the flows according to their priority. The less slots it requires, the higher priority it has, and the earlier it will be scheduled. Algorithm \ref{Algorithm:2} summarizes the scheduling subproblem algorithm based on the quality of service.
	
	\subsection{Complexity Analysis}
	
	The complexity of the sum scheduling flows maximization algorithm is not only related to the complexity of the power allocation subproblem and the phase shift optimization subproblem, but also related to the complexity of scheduling subproblem. For the former one, for each RIS element $l$, the local search algorithm keeps the remaining phase shifts unchanged, selects the best one among $2^e$  phase shifts, and updates a value for $\phi_l$. Since there are L elements, the complexity of this part is $\mathcal{O}(N\times2^e)$. Once updating an element, we calculating the best beamforming vector for $\phi_l$, its computing complexity is derived from formular $(28)$ and $(29)$, and equals to $\mathcal{O}(M^2\times L^6)$. So the computing complexity of the former part is $\mathcal{O}(M^2\times L^6\times N×2^e)$. For the scheduling part, we need to the above operations for the all flows, so the complexity of this part is $\mathcal{O}(F)$. Therefore, the complexity of the proposed algorithm can be given by $\mathcal{O}(FNM^2 L^6 2^e)$.

	\section{Simulation Results} \label{section: Simulation Results}
	In this section, we evaluate the performance of our RIS-assisted wireless resource scheduling scheme. By comparing with other benchmark algorithms, we get numerical results under different representative parameters, verify the effectiveness of our proposed scheme, and study the impact of different parameters on the system performance.
	
	\subsection{Simulation Setup}
	We consider a mmWave HSR system. The train has eight carriages with a total length of 200 meters. There are 24 MRs evenly deployed on the rooftop of the train. So, each carriage is equipped with 1 or 2 MRs, and the distance between two adjacent MRs is 8.33m. The vertical distance between the BS and the track is 75m. The shortest distance between the BS and the MRs is 75m, and the longest distance is 213.6m. In addition, there is an eavesdropper randomly distributed  around the MRs on the rooftop. All communication links in the system are transmitted within the communication range allowed by the transceiver. The number of flows need to be scheduled between the BS and MRs is less than the number of MRs. The QoS of each flow is evenly distributed between 5Mbps$\sim$100Mbps. The other parameters set in the simulation are shown in Table \ref{table1}. 	

		\begin{figure}[!t]
		\centering
		\includegraphics[width=3.5in,height=1.5in]{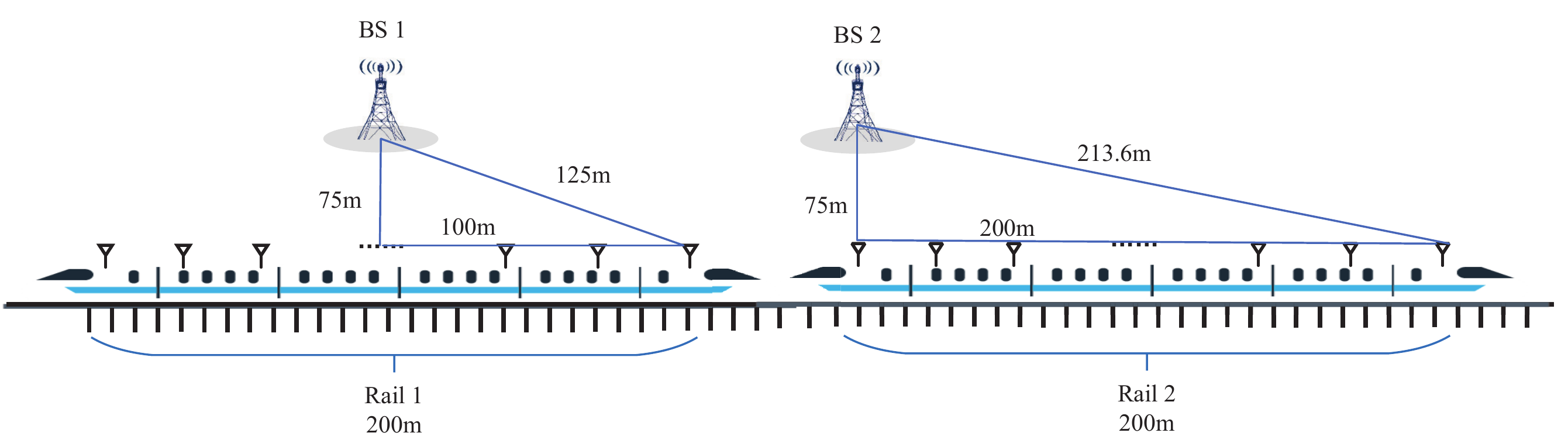}
		\caption{Distance between the BS and MRs.}
		\label{Fig:2}
		\label{fig_1}
	\end{figure}

	\begin{table}[!t]
		\caption{Simulation Parameters\label{tab:table1}}
		\label{table1}
		\centering
		\begin{tabular}{lll}
			\hline
			Parameter & Symbol & Value\\
			\hline
			Carrier frequency & $f$ & 28GHz\\
			Bandwidth & $W$ & 100MHz\\
			Max distance between BS and MRs & $d_{n,1}$ & 213.6m\\
			Min distance between BS and MRs & $d_{n,1}$ & 75m\\
			Distance between BS and RIS & $d_{R,2}$ & 50m\\
			Distance between RIS and MRs & $d_{n,3}$ & 30m\\
			Path loss intercept  & $C$ & -61.3dB\\
			Path loss exponent in the LoS case & $\alpha_1$ & 2.5\\
			Path loss exponent in the NLoS case & $\alpha_2$ & 3.6\\
			Number of BS transmit antennas & $M$ & 4\\
			Number of RIS elements & $L$ & 30\\
			Noise power spectral density & $N_0/W$ & -134dBm/MHz\\
			Maximum transmission power &$P_{max}$ & 23dBm\\
			Beacon period &$T_s$ & 850us\\
			slot time & $T$ & 18us\\
			The number of slots & $K$ & 2000\\
			Rician factor & $\beta_l$ & 4\\
			Number of flows & $F$ & 15\\
			\hline		
		\end{tabular}
	\end{table}

	\subsubsection*{\textup{During the evaluation study phase, we compare the following indicators to evaluate the performance of the \textbf{Proposed-algorithm} and the benchmark algorithms}}
	\begin{enumerate}
		\item{\textbf{The number of scheduled flows:} the number of schedulable flows that meet their QoS requirements. If the QoS requirements are not met, the corresponding scheduled flows are not counted as completed flows.}
		\item{\textbf{System security capacity:} the link secret capacity of the legal MRs.This metric is the difference between the throughput of the legitimate user link and that of the eavesdropper link in the network.}
		\item{\textbf{Total scheduled time slots:} the total number of time slots required by all scheduled flows. The actual value of this metric must be less than the total number of slots in the frame.}
	\end{enumerate}
	
	\subsubsection*{\textup{In order to show the system performance of the \textbf{Proposed-algorithm}, we compare it with the following algorithms}}
	\begin{itemize}
		\item{\textbf{Without-RIS:} this scheme does not use RIS to reflect the signal, and so the receiver can only receive the signal through the direct link. The same power allocation algorithm is used to reduce interference and eavesdropping.}
		\item{\textbf{Manifold optimize:} this algorithm considers continuous RIS phase shifts. A manifold optimization algorithm is then used to iteratively optimize the power allocation of $P_{max}$ and the continuous phase shift matrix.}
		\item{\textbf{Random Phase Shift (RPS):} the algorithm randomly selects a feasible phase shift for each RIS element and keeps these phase shifts unchanged. Then, the optimal beamforming vector at the transmitter is obtained by using the same power allocation algorithm as the proposed RIS-assisted algorithm. Then the goal of maximizing the number of schedulable flows is discussed.}
		\item{\textbf{Average Power Transmission (APT):} the scheme allocates the same transmit power to all transmit antennas, and then uses the local search algorithm to find the RIS phase shift that maximizes the security capacity of each link. Then it discusses the goal of maximizing the number of schedulable flows.}
	\end{itemize}

	\subsection{Performance Analysis}

\begin{figure}[!t]
			\centering
			\includegraphics[width=3.2in,height=2.5in]{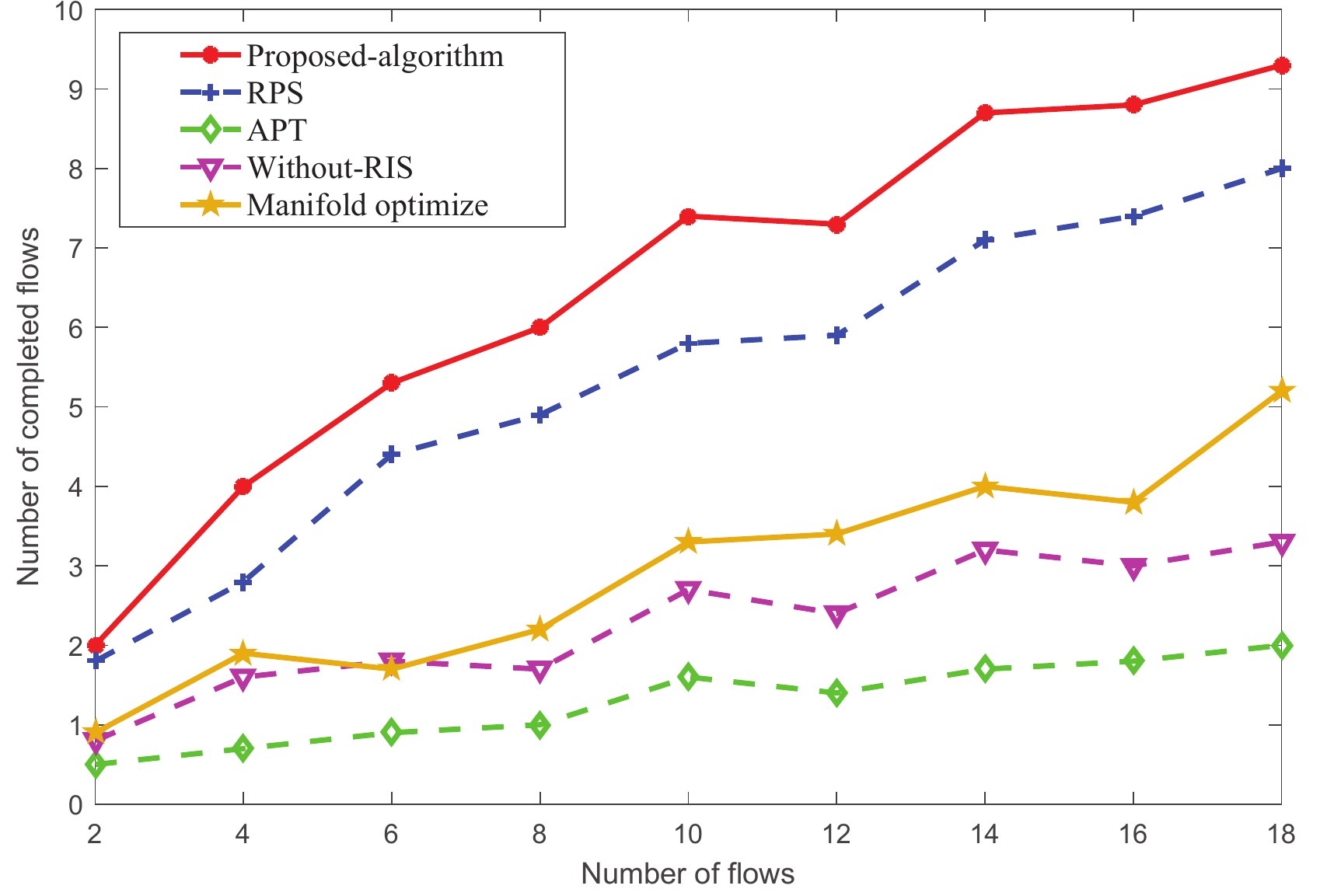}
			\caption{Number of completed flows vs different numbers of requested flows.}
			\label{Fig:3}
			\label{fig_1}
		\end{figure}

	In Fig. \ref{Fig:3}, we set the number of RIS elements $L = 30$, the number of quantization bits $e = 3$, and draw the curves of the number of schedulable flows of these four schemes as the number of requested flows changes from 2 to 18. It is observed that the trend of all five schemes increases as the number of requested flows increases, which indicates that the more requested flows in the HSR network, the more flows can be scheduled on a limited scale. The number of the flows that the proposed-agorithm can schedule is larger than the other four schemes. In particular, it is $185\%$ larger than that of without RIS. It shows that the system capacity of communication links can be improved by RIS's reflection, and more flows can be scheduled within the same time slots. The APT scheme achieves much less flows than the other four schemes. This is because the average transmit power does not take into account the channel capacity of the eavesdropper as low as possible. In the case of confidentiality, the number of schedulable flows is correspondingly small. In addition, compared with RPS, the proposed RIS-assisted scheduling scheme has more schedulable flows, which shows that choosing the appropriate RIS reflecting element phase shift can also increase the channel capacity. It is worth mentioning that, considering continuous RIS phase shifts, the performance optimized with the manifold optimization algorithm is shown by the yellow line in Fig. \ref{Fig:3}. Although it successfully schedules more flows than the Without-RIS algorithm, it is much less than our proposed-algorithm. Because in this case, although the RIS can provide a certain reflection gain, the manifold optimization only obtains a suboptimal solution of the constrained variables. While our algorithm obtains the optimal closed-form solution. Moreover, manifold optimization takes many iterations to converge to the suboptimal solution, which is more complex than our proposed-algorithm. In short, when the number of requested flows is 18, the performance of the proposed RIS-assisted algorithm is $1.2288$, $5.1221$, $2.8585$ and $1.7520$ times that of RPS, APT, without-RIS and manifold optimization, respectively. Therefore, compared with the other four schemes, the proposed-algorithm has the best performance in the number of completed flows.

		\begin{figure*}[t]
		\centering
		\subfloat[]{\includegraphics[width=3.2in,height=2.5in]{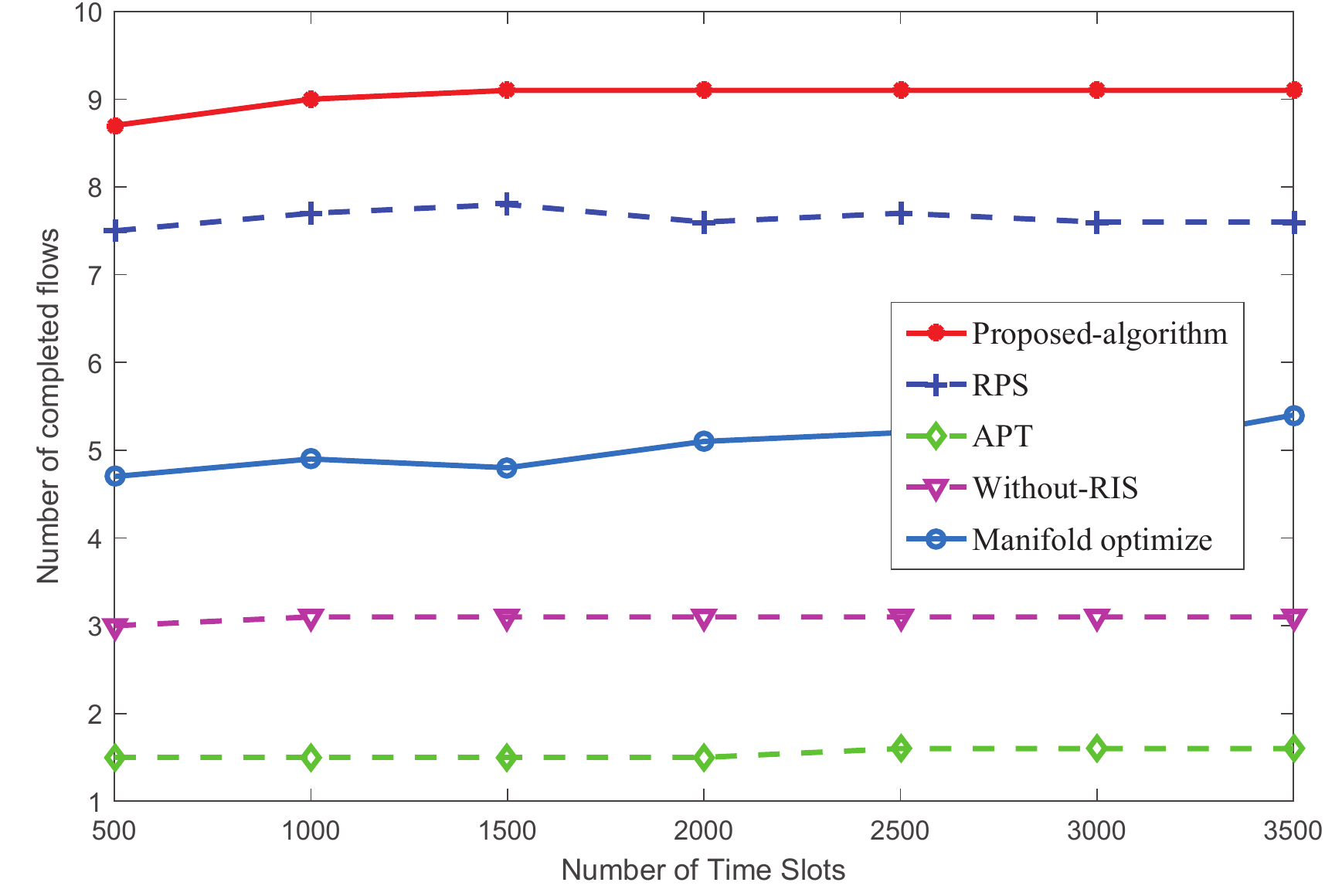}%
			\label{fig_first_case}}
		\hfil
		\subfloat[]{\includegraphics[width=3.2in,height=2.5in]{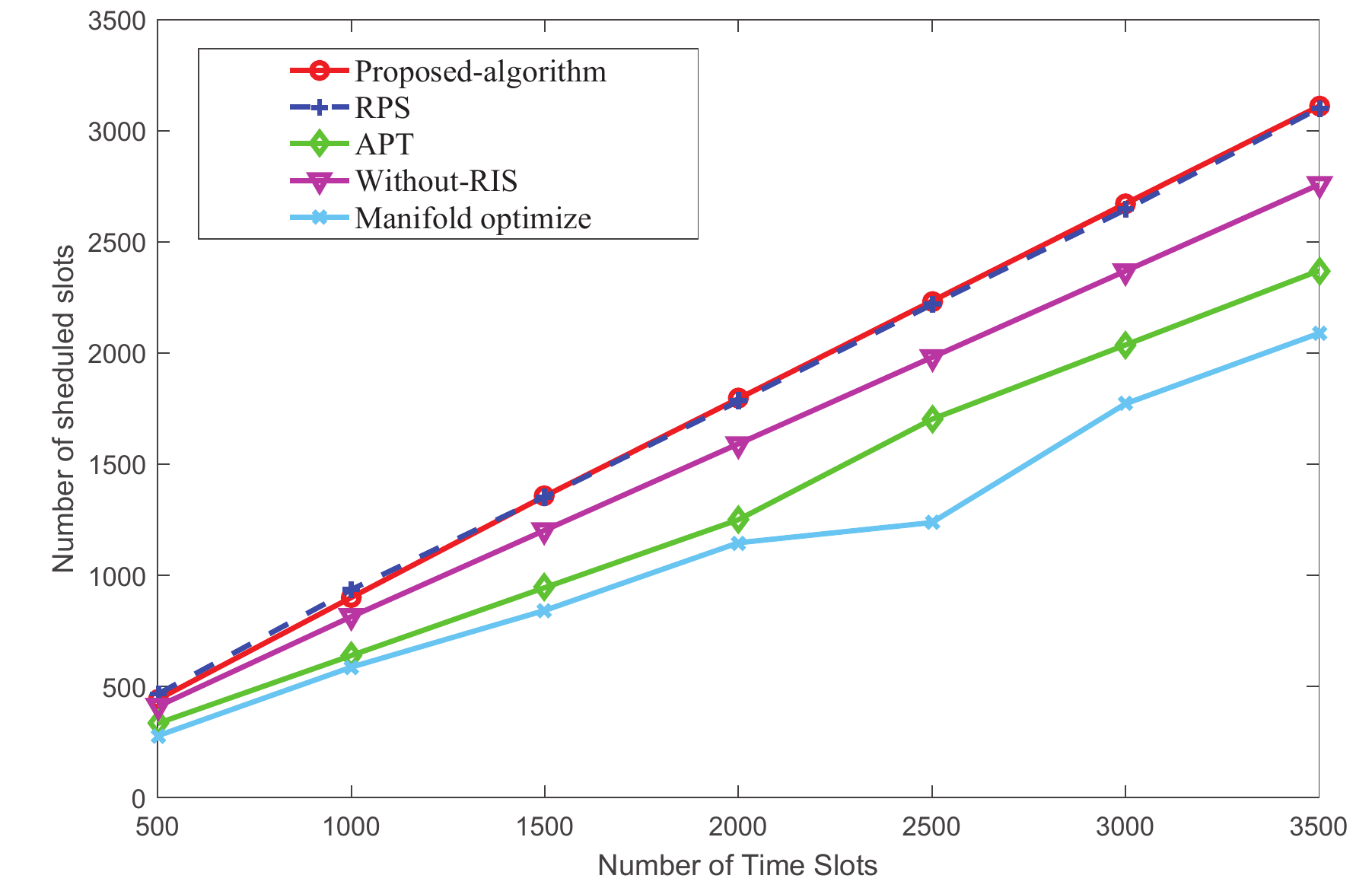}%
			\label{fig_second_case}}
		\caption{(a) Number of completed flows vs Number of TSs. (b) Number of scheduled slots vs Number of TSs.}
		\label{Fig:4}
		\label{fig_sim}
	\end{figure*}
			
\begin{figure}[t]
	\centering
	\includegraphics[width=3.2in,height=2.5in]{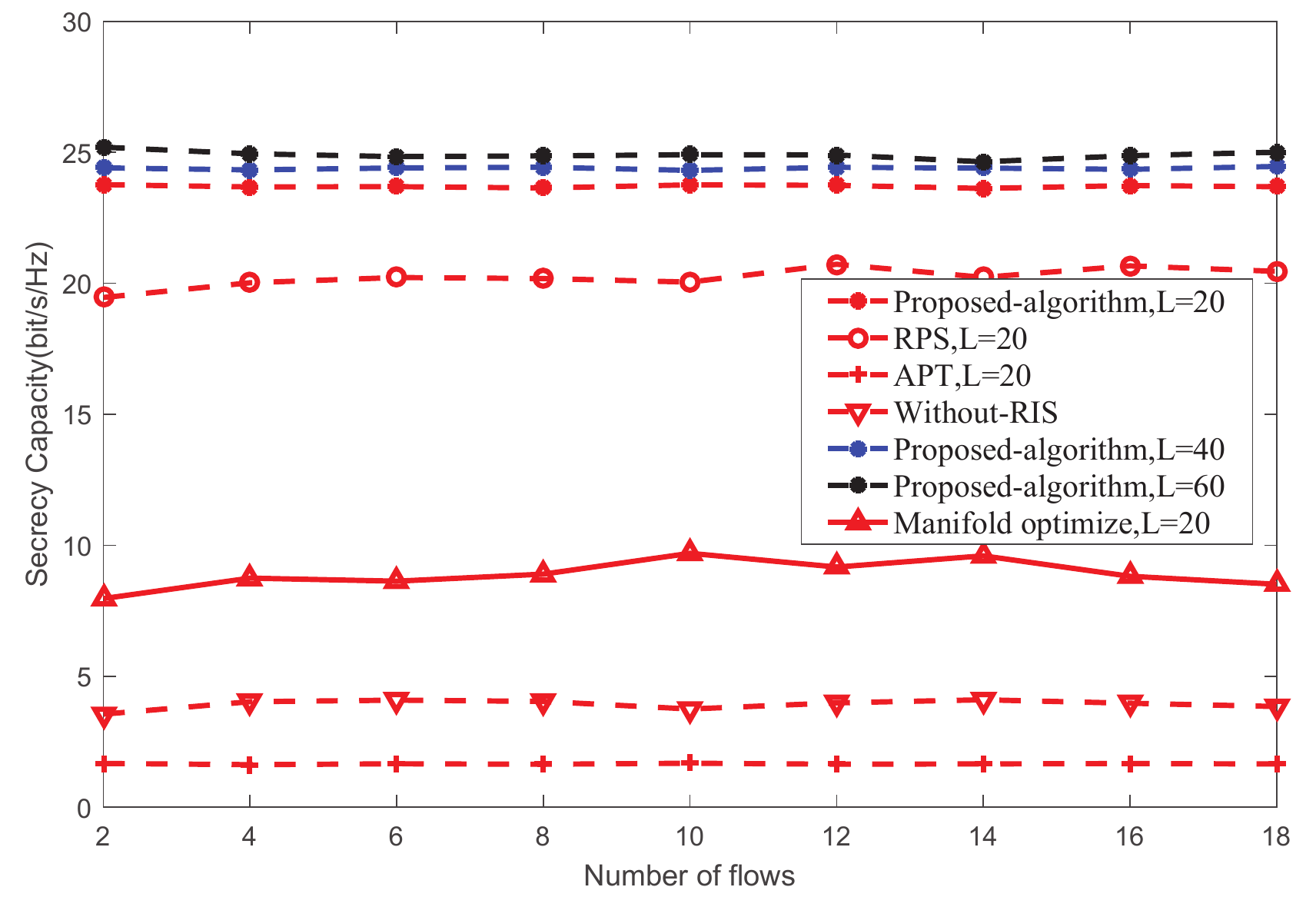}
	\caption{Secrecy capacity vs the number of RIS elements L.}
	\label{Fig:5}
	\label{fig_1}
\end{figure}

\begin{figure}[t]
	\centering
	\includegraphics[width=3.2in,height=2.5in]{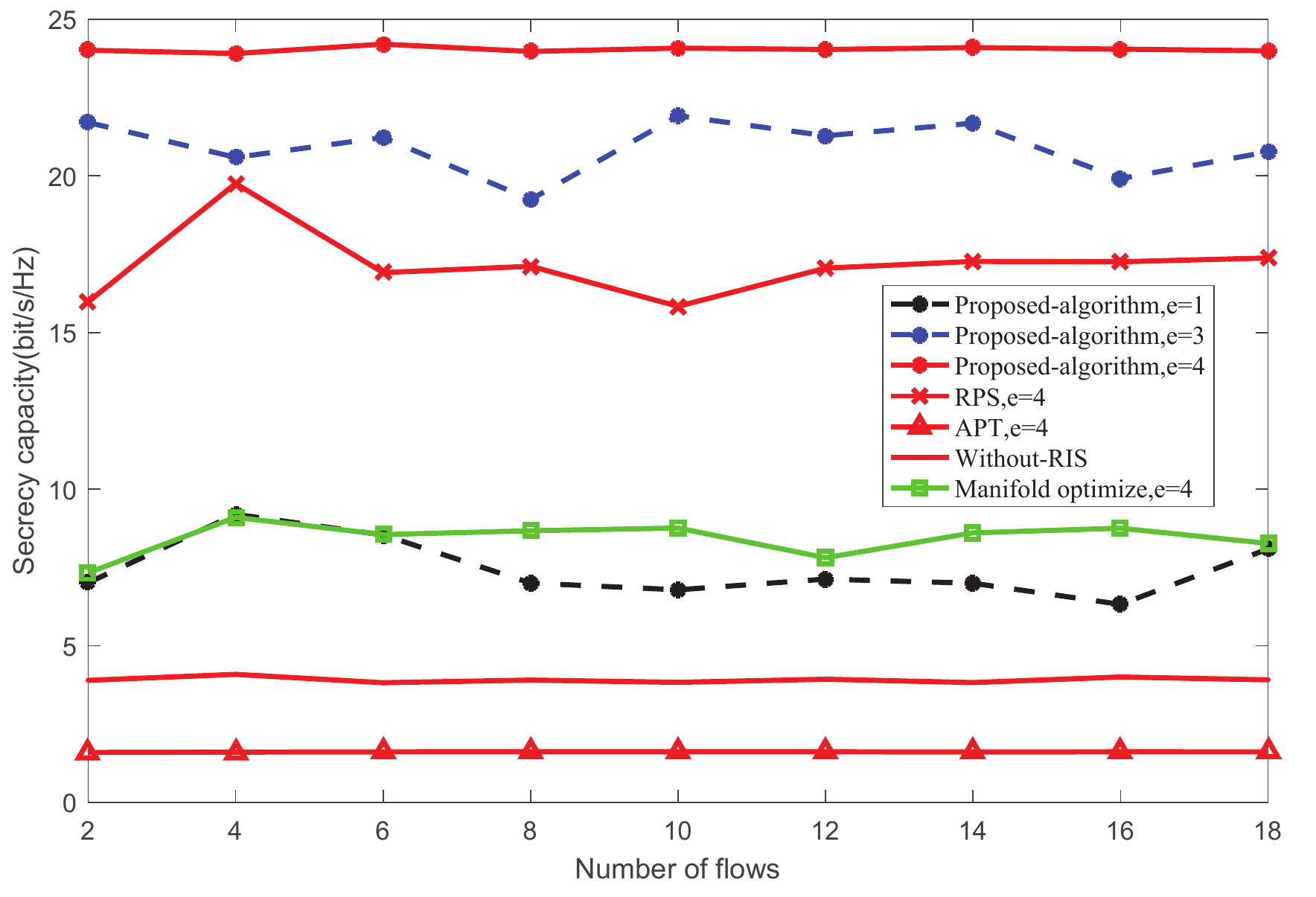}
	\caption{Secrecy capacity vs the number of quantization bits.}
	\label{Fig:6}
	\label{fig_1}
\end{figure}
	
	In Fig. \ref{Fig:4}, we also set $L = 30$, $e = 3$, and the number of requested flows is 18. We plot the number of schedulable flows and the number of time slots used under different time slots per frame. It can be seen that when the number of slots increases, the numbers of schedulable flows of the five schemes vary little and are almost flat. Therefore, the number of time slots does not need to be increased indefinitely. High performance can be achieved by setting an appropriate number of time slots according to the number of requested flows and corresponding QoS requirements. The number of flows that can be scheduled in the proposed RIS-assisted scheduling scheme is $1.2293$, $5.3819$, $2.6575$ and $1.6932$ times that of RPS, APT, without-RIS and manifold optimization schemes, respectively. The number of time slots used is close to that of the other three schemes. It indicates that the proposed RIS-assisted scheduling algorithm has better performance in transmission efficiency.

	In Fig. \ref{Fig:5}, we set $e=3$ and $F=18$. We draw the curve of channel security capacity as the number of requested flows changes when $L=20$, $L=40$ and $L=60$. It can be seen from the figure that the number of schedulable flows of the proposed RIS-assisted scheduling scheme is the largest than the other schemes. Moreover, as the value of $L$ increases, the amount of secrecy capacity alse increases gradually. This illustrates that an appropriate increase in the number of RIS reflective elements can  effectively enhance channel quality.

	In Fig. \ref{Fig:6}, we set $L=30$ and plot the change of channel secrecy capacity as the number of requested flows changes, with the phase shift quantization number of RIS reflection elements $e$ setted to 1, 3 and 4. It can be seen from the figure that the larger the quantization number of RIS reflection phase shift is, the larger system secrecy capacity can be obtained. This is because with the increase of $e$, the RIS elements can adjust the phase more accurately, and a more optimal phase shift can be found through the local search algorithm, so as to obtain a better secrecy capacity.
	
\begin{figure}[t]
		\centering
		\includegraphics[width=3.2in,height=2.4in]{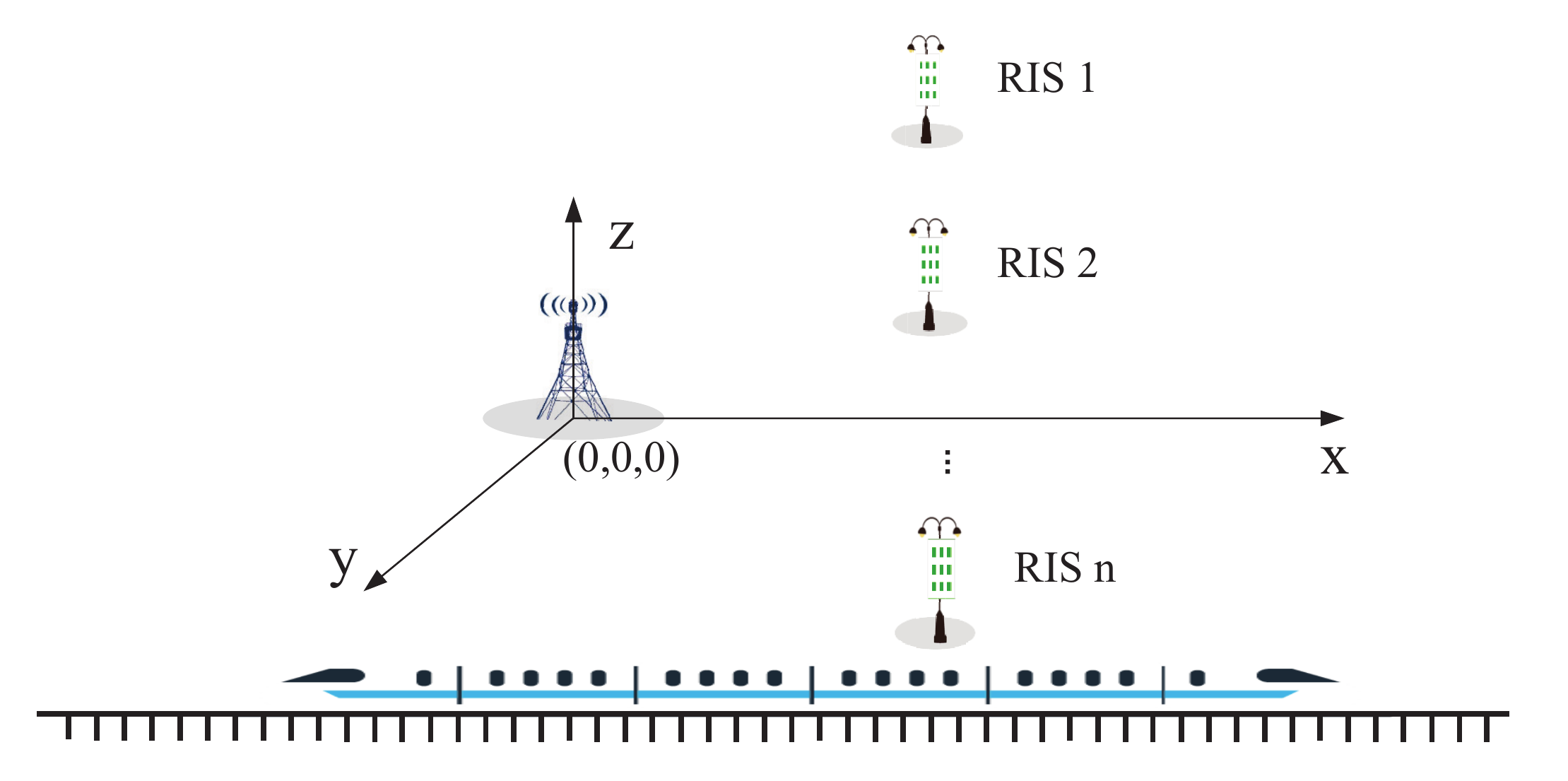}
		\caption{The three-dimensional coordinate system.}
		\label{Fig:7}
		\label{fig_1}
	\end{figure}

\begin{figure}[t]
		\centering
		\includegraphics[width=3.2in,height=2.5in]{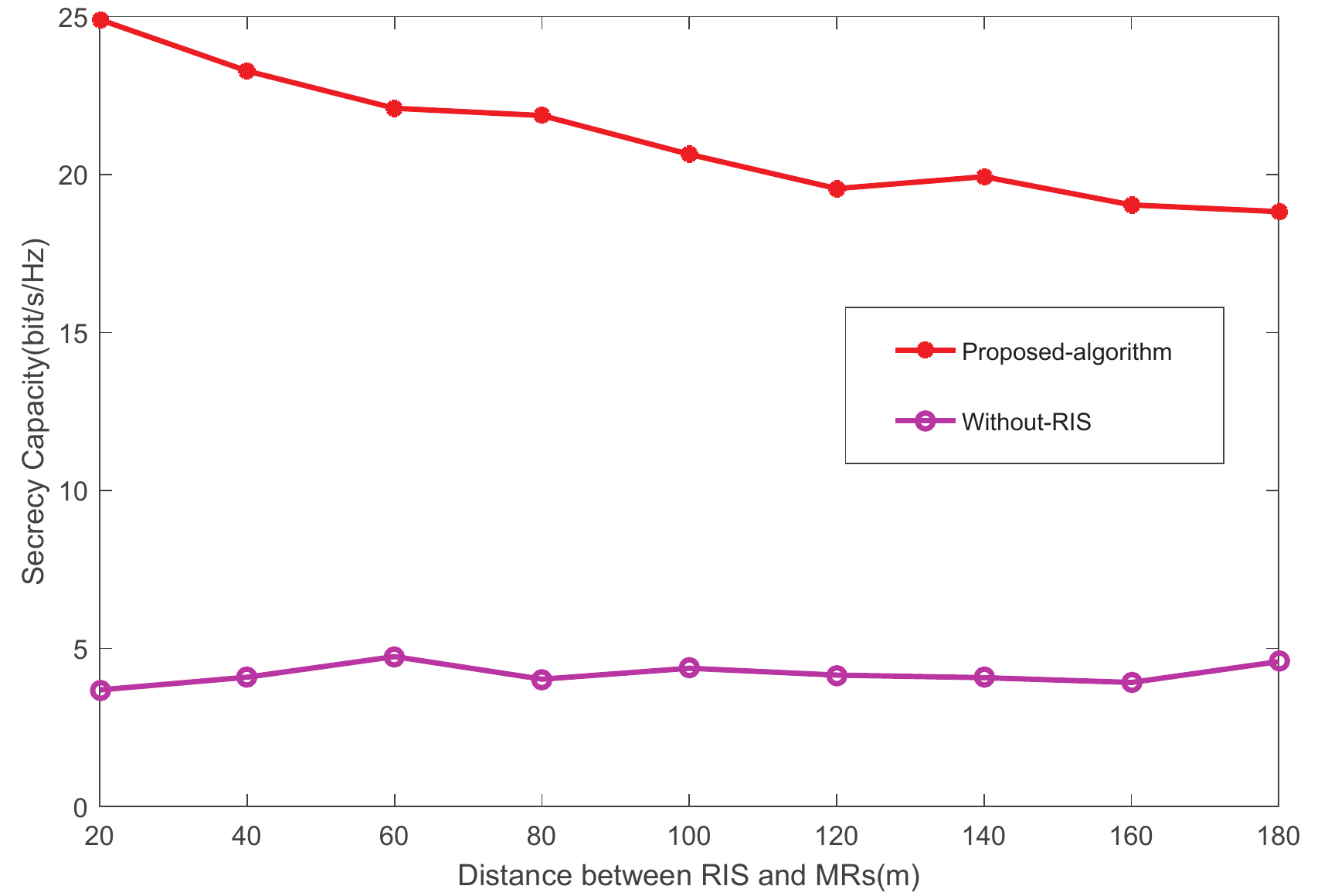}
		\caption{Secrecy capacity vs the distance between RIS and MRs.}
		\label{Fig:8}
		\label{fig_1}
	\end{figure}

\begin{figure}[t]
		\centering
		\includegraphics[width=3.2in,height=2.4in]{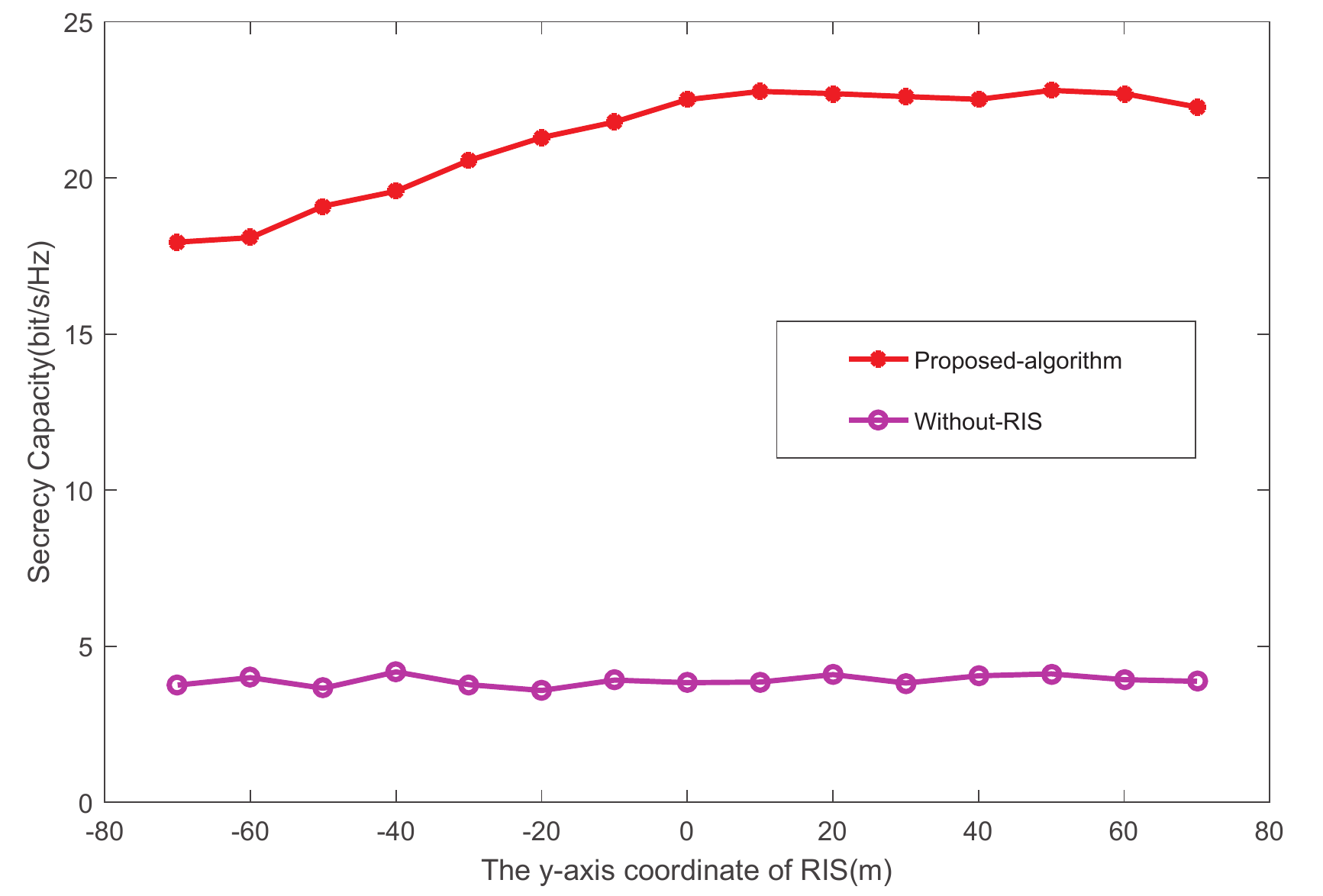}
		\caption{Secrecy capacity vs the position of RIS.}
		\label{Fig:9}
		\label{fig_1}
	\end{figure}

	In Fig. \ref{Fig:8} and Fig. \ref{Fig:9}, we set $L=30$ and $e=4$, and we plot the impact of where RIS is deployed on system secrecy capacity. Specifically, in Fig. \ref{Fig:8} we only change the distance between RIS and MRs to observe the change in performance. It can be seen that as the distance of RIS-MRs increases, the security capacity decreases gradually. This is because as the reflection distance increases, the double fading effect of the RIS also increases gradually. This makes the auxiliary function of RIS weakened, so that the security capacity is gradually reduced. In Fig. \ref{Fig:8} we change the deployment location of RIS in another way. As shown in Figure. \ref{Fig:7}, we establish a three-dimensional coordinate system centered on the position of the BS. Fix other parameters unchanged, only gradually increase $y_r$ from a negative value to a positive value. In this way, the distance changes of BS-RIS and RIS-MRs can be reflected simultaneously. Fig. \ref{Fig:9} shows that as  $y_r$ increases from a negative value to 0 and then to a positive value, the security capacity firstly increases gradually and then tends to be stable. This is because as the distance of BS-RIS-MRs gradually decreases, the double fading effect of RIS gradually weakens, and the security capacity of the system gradually increases. But when the RIS gradually approaches the MRs from the vicinity of the BS, the security capacity no longer changes significantly. This is because the distance change of BS-RIS-MRs does not change much, and on the other hand, the improvement of system performance caused by RIS gradually tends to be saturated. In general, even if RIS is a little far from the BS and MRs, the performance is slightly impaired, but compared with the situation without RIS, it still greatly improves the security capacity.
	
\begin{figure}[h]
		\centering
		\includegraphics[width=3.2in,height=2.5in]{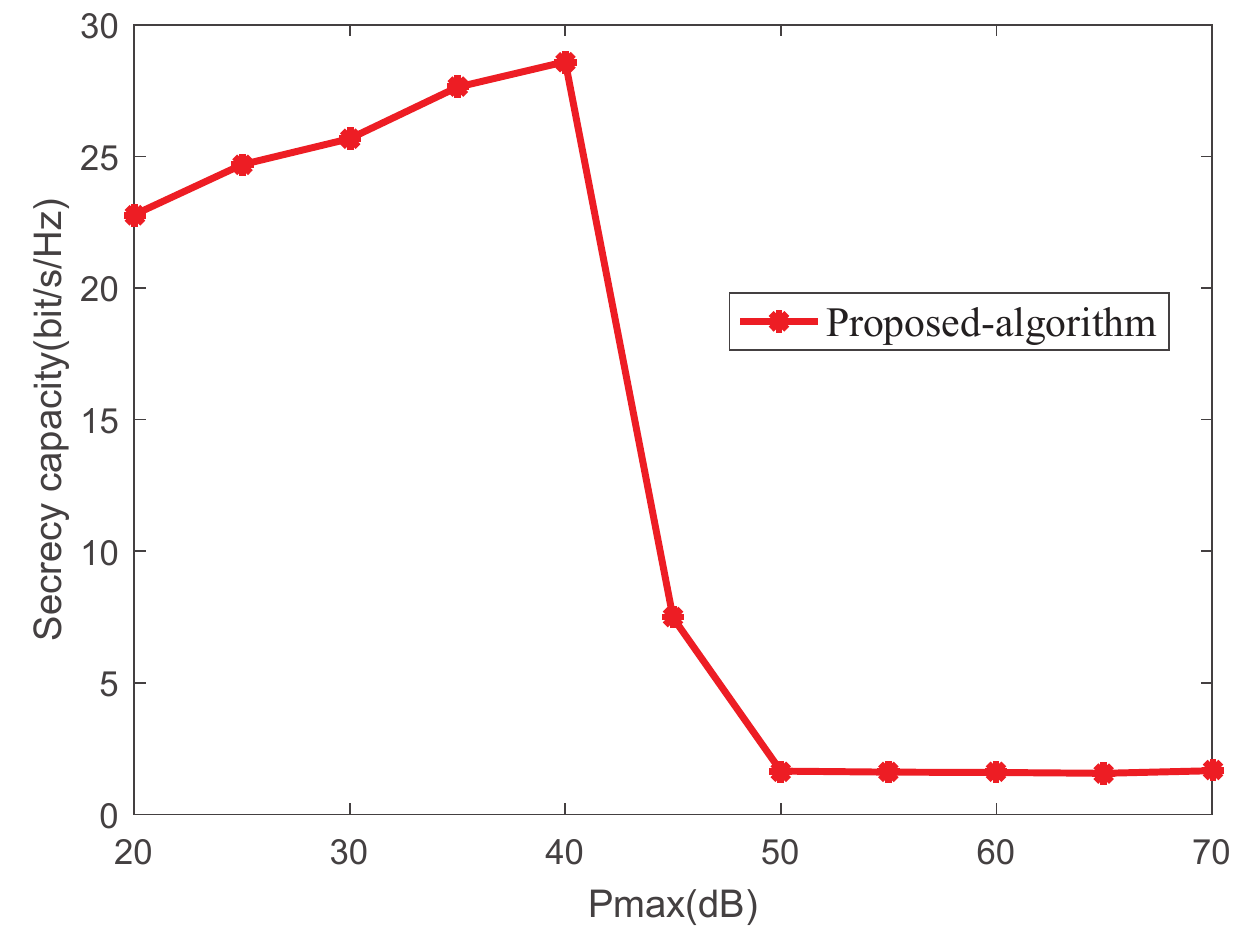}
		\caption{Secrecy capacity vs transmit power Pmax.}
		\label{Fig:10}
		\label{fig_1}
	\end{figure}

	In Fig. \ref{Fig:10}, we set $L=30$ and $e=4$ to plot the influence of the BS transmitting power on the system secrecy capacity. As can be seen from Fig. \ref{Fig:10}, when the transmitting power is small, the system secrecy capacity of the RIS-assisted scheduling scheme increases gradually as the increase of the BS transmitting power, which indicates that the increase of transmitting power can effectively improve the system secrecy capacity. However, when $P_{max}$ increases to a certain extent, as shown in Fig. \ref{Fig:10} is about 40dB, the system secrecy capacity of the proposed RIS-assisted  scheduling scheme drops sharply. This is because when $P_{max}$ increases to a certain extent, although the system capacity of the legitimate receiver increases, the system capacity of the eavesdropper also increases to the point where it is difficult to suppress, leading to the secrecy capacity very small. Therefore, the transmission power is not the bigger the better, it should be reasonable setted to maximize the system secrecy capacity.
	
	\section{Conclusion} \label{section: Conclusion}
	In this paper, we have focused on the problem of scheduling flows with diverse QoS requirements in the mmWave HSR communication scenario. Under the constraints of maximum transmit power, discrete phase shift and minimum secrecy capacity, we joint the power allocation design, the RIS' phase shift optimization and scheduling order problem to maximize the number of successfully scheduled flows. Extensive simulations have showed that the proposed RIS-assisted algorithm outperformed the other three baseline schemes on the number of completed flows and achievable secrecy capacity. In the future work, we will consider RIS-assisted multi-cell HSR communication to reduce the number of handovers and improve the handover success rate. We will also consider adopting the FDMA scheme, using RIS to enhance the useful signal, and achieve the purpose of reducing interference at the same time, so as to maximize the utilization of frequency resources.

	\begin{IEEEbiography}[{\includegraphics[width=1in,height=1.25in,clip,keepaspectratio]{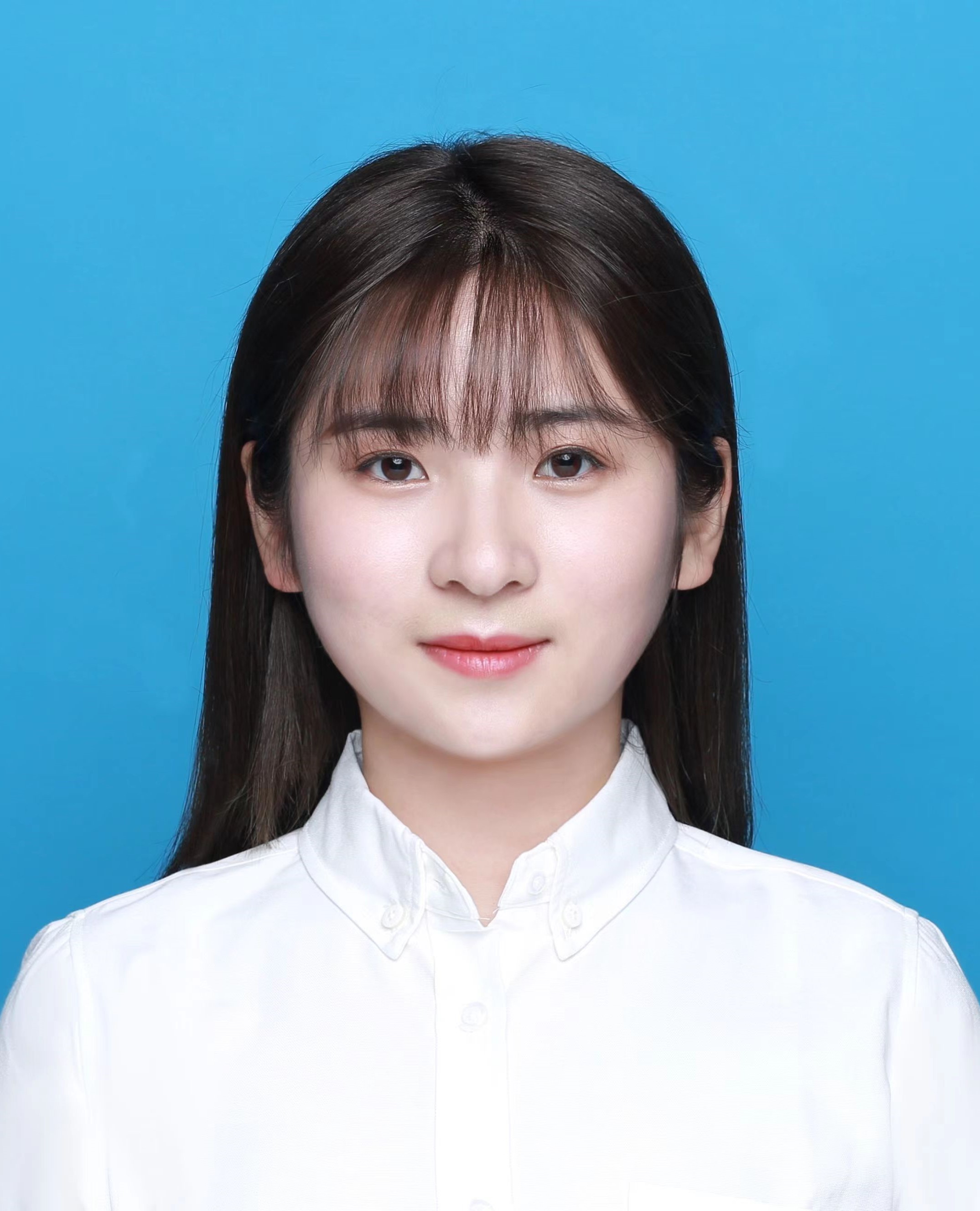}}]{Panpan Li}
		was born in Henan, China, in 1997. She received the B.S. degree in communication engineering from North China Electric Power University, Baoding, China, in 2019. She is currently pursuing the Ph.D. degree with the State Key Laboratory of Rail Traffic Control and Safety, Beijing Jiaotong University, Beijing, China. Her current research includes millimeter-wave wireless communications, reconfigurable intelligent surfaces, convex optimization and wireless resource allocation.
	\end{IEEEbiography}

	\begin{IEEEbiography}[{\includegraphics[width=1in,height=1.25in,clip,keepaspectratio]{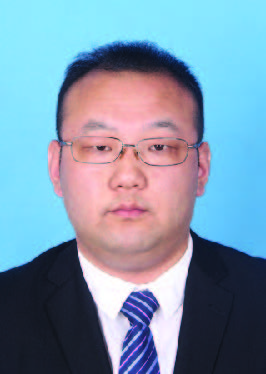}}]{Yong Niu}
		(M’17) received the B.E. degree in Electrical Engineering from Beijing Jiaotong University, China, in 2011, and the Ph.D. degree in Electronic Engineering from Tsinghua University, Beijing, China, in 2016.
		
		From 2014 to 2015, he was a Visiting Scholar with	the University of Florida, Gainesville, FL, USA. He is currently an Associate Professor with the State Key Laboratory of Rail Traffic Control and Safety, Beijing Jiaotong University. His research interests are in the areas of networking and communications, including millimeter wave communications, device-to-device communication, medium access control, and software-defined networks. He received the Ph.D. National Scholarship of China in 2015, the Outstanding Ph.D. Graduates and Outstanding Doctoral Thesis of Tsinghua University in 2016, the Outstanding Ph.D. Graduates of Beijing in 2016, and the Outstanding Doctorate Dissertation Award from the Chinese Institute of Electronics in 2017. He has served as Technical Program Committee member for IWCMC 2017, VTC2018-Spring, IWCMC 2018, INFOCOM 2018, and ICC 2018. He was the Session Chair for IWCMC 2017. He was the recipient of the 2018 International Union of Radio Science Young Scientist Award.
	\end{IEEEbiography}

	\begin{IEEEbiography}[{\includegraphics[width=1in,height=1.25in,clip,keepaspectratio]{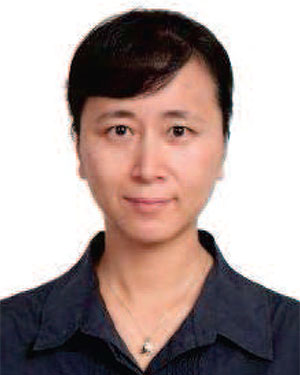}}]{Hao Wu}
		received the Ph.D. degree in information and
		communication engineering from the Harbin Institute
		of Technology, Harbin, China, in 2000. She is currently
		a Full Professor with the State Key Lab of Rail
		Traffic Control and Safety, Beijing Jiaotong University,
		Beijing, China. She has authored or coauthored
		more than 100 papers in international journals and
		conferences.Her research interests include intelligent
		transportation systems, security and QoS issues in
		wireless networks (VANETs,MANETs, and WSNs),
		wireless communications, and Internet of Things. She
		is a Reviewer of major conferences and journals in wireless networks and
		security for IEEE.
	\end{IEEEbiography}

	\begin{IEEEbiography}[{\includegraphics[width=1in,height=1.25in,clip,keepaspectratio]{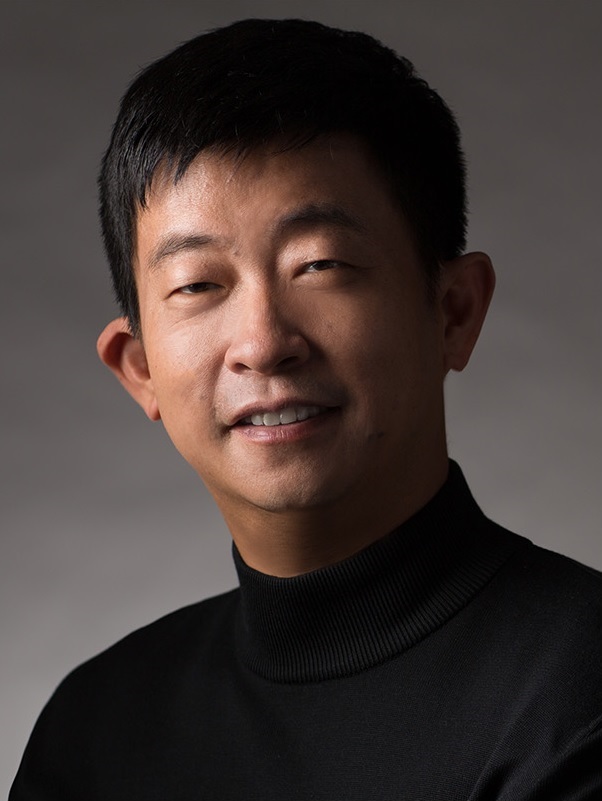}}]{Zhu Han}
		 (S’01–M’04-SM’09-F’14) received the B.S. degree in electronic engineering from Tsinghua University, in 1997, and the M.S. and Ph.D. degrees in electrical and computer engineering from the University of Maryland, College Park, in 1999 and 2003, respectively. 
		
		From 2000 to 2002, he was an R\&D Engineer of JDSU, Germantown, Maryland. From 2003 to 2006, he was a Research Associate at the University of Maryland. From 2006 to 2008, he was an assistant professor at Boise State University, Idaho. Currently, he is a John and Rebecca Moores Professor in the Electrical and Computer Engineering Department as well as in the Computer Science Department at the University of Houston, Texas. His research interests include wireless resource allocation and management, wireless communications and networking, game theory, big data analysis, security, and smart grid.  Dr. Han received an NSF Career Award in 2010, the Fred W. Ellersick Prize of the IEEE Communication Society in 2011, the EURASIP Best Paper Award for the Journal on Advances in Signal Processing in 2015, IEEE Leonard G. Abraham Prize in the field of Communications Systems (best paper award in IEEE JSAC) in 2016, and several best paper awards in IEEE conferences. Dr. Han was an IEEE Communications Society Distinguished Lecturer from 2015-2018, AAAS fellow since 2019, and ACM distinguished Member since 2019. Dr. Han is a 1\% highly cited researcher since 2017 according to Web of Science. Dr. Han is also the winner of the 2021 IEEE Kiyo Tomiyasu Award, for outstanding early to mid-career contributions to technologies holding the promise of innovative applications, with the following citation: ``for contributions to game theory and distributed management of autonomous communication networks."						
	\end{IEEEbiography}

	\begin{IEEEbiography}[{\includegraphics[width=1in,height=1.25in,clip,keepaspectratio]{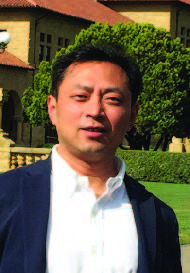}}]{Bo Ai}
		Bo Ai received the M.S. and Ph.D. degrees from Xidian University, China. He studies as a Post-Doctoral Student at Tsinghua University. He was a Visiting Professor with the Electrical Engineering Department, Stanford University, in 2015. He is currently with Beijing Jiaotong University as a Full Professor and a Ph.D. Candidate Advisor. He is the Deputy Director of the State Key Lab of Rail Traffic Control and Safety and the Deputy Director of the International Joint Research Center. He is one of the
		main people responsible for the Beijing Urban Rail Operation Control System, International Science and Technology Cooperation Base. He is also a Member, of the Innovative Engineering Based jointly granted by the Chinese Ministry of Education and the State Administration of Foreign Experts Affairs. He was honored with the Excellent Postdoctoral Research Fellow by Tsinghua University in 2007.
		
		He has authored/co-authored eight books and published over 300 academic research papers in his research area. He holds 26 invention patents. He has been the research team leader for 26 national projects. His interests include the research and applications of channel measurement and channel modeling, dedicated mobile communications for rail traffic systems. He has been notified by the Council of Canadian Academies that, based on Scopus database, he has been listed as one of the Top 1\% authors in his field all over the world. He has also been feature interviewed by the IET Electronics Letters. He has received some important scientific research prizes.
		
		Dr. Ai is a fellow of the Institution of Engineering and Technology. He is an Editorial Committee Member of the Wireless Personal Communications journal. He has received many awards, such as the Outstanding Youth Foundation from the National Natural Science Foundation of China, the Qiushi Outstanding Youth Award by the Hong Kong Qiushi Foundation, the New Century Talents by the Chinese Ministry of Education, the Zhan Tianyou
		Railway Science and Technology Award by the Chinese Ministry of Railways, and the Science and Technology New Star by the Beijing Municipal Science and Technology Commission. He was a co-chair or a session/track chair for  many international conferences. He is an IEEE VTS Beijing Chapter Vice Chair and an IEEE BTS Xi’an Chapter Chair. He is the IEEE VTS Distinguished Lecturer. He is an Editor of the IEEE TRANSACTIONS ON CONSUMER ELECTRONICS. He is the Lead Guest Editor of Special Issues of the IEEE TRANSACTIONS ON VEHICULAR TECHNOLOGY, the IEEE ANTENNAS AND WIRELESS PROPAGATION LETTERS, and the International Journal of Antennas and Propagation.
	\end{IEEEbiography}

	\begin{IEEEbiography}[{\includegraphics[width=1in,height=1.25in,clip,keepaspectratio]{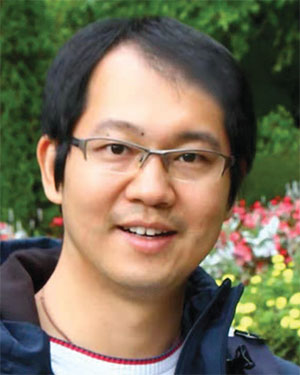}}]{Ning Wang}
		(Member, IEEE) received the B.E. degree in communication engineering from Tianjin University, Tianjin, China, in 2004, the M.A.Sc. degree in electrical engineering from The University of British Columbia, Vancouver, BC, Canada, in 2010, and the Ph.D. degree in electrical engineering from the University of Victoria, Victoria, BC, Canada, in 2013. From 2004 to 2008, he was with the China Information Technology Design and Consulting Institute, as a Mobile Communication System Engineer, specializing in planning and design of commercial mobile communication networks, network traffic analysis, and radio network optimization. From 2013 to 2015, he was a Postdoctoral Research Fellow with the Department of Electrical and Computer Engineering, The University of British Columbia. Since 2015, he has been with the School of Information Engineering, Zhengzhou University, Zhengzhou, China, where he is currently an Associate Professor. He also holds adjunct appointments with the Department of Electrical and Computer Engineering, McMaster University, Hamilton, ON, Canada, and the Department of Electrical and Computer Engineering, University of Victoria. His research interests include resource allocation and security designs of future cellular networks, channel modeling for wireless communications, statistical signal processing, and cooperative wireless communications. He was on the Technical Program Committees of international conferences, including the IEEE GLOBECOM, IEEE ICC, IEEE WCNC, and CyberC. He was on the Finalist of the Governor Generals Gold Medal for Outstanding Graduating Doctoral Student with the University of Victoria in 2013.
	\end{IEEEbiography}

	\begin{IEEEbiography}[{\includegraphics[width=1in,height=1.25in,clip,keepaspectratio]{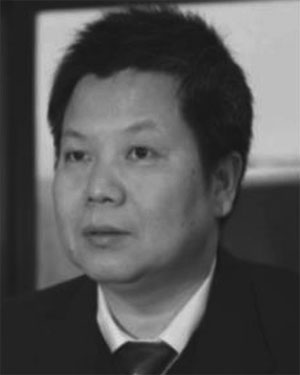}}]{Zhangdui Zhong}
		(SM’16) received the B.E. and M.S. degrees from Beijing Jiaotong University, Beijing, China, in 1983 and 1988, respectively. He is currently a Professor and an Advisor of Ph.D. candidates with Beijing Jiaotong University, where he is also currently a Chief Scientist of the State Key Laboratory of Rail Traffic Control and Safety. He is also the Director of the Innovative Research Team, Ministry of Education, Beijing, and a Chief Scientist of the Ministry of Railways, Beijing.
		
		He is also an Executive Council Member of the Radio Association of China, Beijing, and a Deputy Director of the Radio Association, Beijing. His interests include wireless communications for railways, control theory and techniques for railways, and GSM-R systems. His research has been widely used in railway engineering, such as the Qinghai-Xizang railway, DatongQinhuangdao Heavy Haul railway, and many high-speed railway lines in China. He has authored or co-authored seven books, five invention patents, and over 200 scientific research papers in his research area. Prof. Zhong was a recipient of the Mao YiSheng Scientific Award of China, Zhan TianYou Railway Honorary Award of China, and Top 10 Science/Technology Achievements Award of Chinese Universities.
	\end{IEEEbiography}
	
\end{document}